\documentclass[journal,10pt,a4paper,twoside]{IEEEtran}

\ifCLASSOPTIONdraftcls
	\newcommand{\figheight}{85mm} 
	
\else
  \newcommand{\figheight}{70mm}
  
\fi

\usepackage{ifthen}

\usepackage{amsmath}
\usepackage{amssymb}
\usepackage{amsfonts}
\usepackage{graphicx}
\usepackage{psfrag}
\usepackage{paralist}
\usepackage{times}
\usepackage{amsbsy}
\usepackage{latexsym,bm}
\usepackage{ifsym}
\usepackage{epsfig}
\usepackage{color}
\usepackage{paralist}
\usepackage{psfrag}
\usepackage{cite}

\usepackage{amsthm}
\newtheorem{thm}{Theorem}

\newtheorem{lem}{Lemma}
\theoremstyle{definition}

\usepackage{flushend}

\usepackage{bm}

\newcommand{\ma}[1]{\bm{{#1}}}         

\newcommand{\compl}{\mathbb{C}}        
\newcommand{\real}{\mathbb{R}}         

\newcommand{\e}{{\rm e}} 
\renewcommand{\j}{\mathrm{j}} 

\newcommand{\opof}[2]{\mathop{{\rm #1}}\left\{#2\right\}}         

\newcommand{\realof}[1]{\opof{Re}{#1}}           
\newcommand{\imagof}[1]{\opof{Im}{#1}}           
\newcommand{\diagof}[1]{\opof{diag}{#1}}         
\newcommand{\expvof}[1]{\opof{\mathbb{E}}{#1}}   

\newcommand{\vecof}[1]{\opof{vec}{#1}}           

\newcommand{\expof}[1]{{\e}^{#1}}                      

\newcommand{\subsel}[1]{#1^{({\rm sel})}}
\newcommand{\subselr}[2]{#1_#2^{({\rm sel})}}

\newcommand{\rd}[1]{#1^{({r})}}
\newcommand{\rdT}[1]{#1^{({r})^\trans}}
\newcommand{\rdH}[1]{#1^{({r})^\herm}}

\newcommand{\rdtil}[1]{\tilde{#1}^{({r})}}

\newcommand{\nc}[1]{#1^{({\rm nc})}}
\newcommand{\ncr}[1]{#1^{({\rm nc})(r)}}
\newcommand{\ncrT}[1]{#1^{({\rm nc})(r)^\trans}}
\newcommand{\ncrH}[1]{#1^{({\rm nc})(r)^\herm}}
\newcommand{\ncrtil}[1]{\tilde{#1}^{({\rm nc})(r)}}
\newcommand{\ncrtilH}[1]{\tilde{#1}^{{({\rm nc})(r)}^\herm}}
\newcommand{\ncconj}[1]{#1^{({\rm nc})^\conj}}
\newcommand{\ncT}[1]{#1^{({\rm nc})^\trans}}
\newcommand{\ncH}[1]{#1^{({\rm nc})^\herm}}
\newcommand{\ncinv}[1]{#1^{({\rm nc})^{-1}}}

\newcommand{\ncfba}[1]{\tilde{#1}^{({\rm nc})}}

\newcommand{\ncfbaH}[1]{\tilde{#1}^{({\rm nc})^\herm}}
\newcommand{\ncfbainv}[1]{\tilde{#1}^{({\rm nc})^{-1}}}

\newcommand{\fba}[1]{#1^{({\rm fba})}}
\newcommand{\fbaT}[1]{#1^{({\rm fba})^\trans}}

\newcommand{\fbainv}[1]{#1^{({\rm fba})^{-1}}}

\newcommand{\normof}[2]{\left\|#1\right\|_{#2}}
\newcommand{\twonorm}[1]{\normof{#1}{2}}                  

\newcommand{\inv}{{-1}}          
\newcommand{\pinv}{+}          
\newcommand{\conj}{*}          
\newcommand{\trans}{{\rm T}}   
\newcommand{\herm}{{\rm H}}    

\usepackage{color}

\title{$R$-Dimensional ESPRIT-Type Algorithms for Strictly Second-Order Non-Circular Sources and Their Performance Analysis}
\author{Jens Steinwandt$^*$,~\IEEEmembership{Student Member,~IEEE},
		    Florian Roemer,~\IEEEmembership{Member,~IEEE},\\ 
		    Martin Haardt,~\IEEEmembership{Senior Member,~IEEE}, and 
		    Giovanni Del Galdo,~\IEEEmembership{Member,~IEEE}

\thanks{
Parts of this paper have been published at the {\em IEEE Int. Conference on Acoustics, Speech, 
and Signal Processing (ICASSP 2004)}, Montreal, Canada, May 2004, and the {\em IEEE Int. 
Conference on Acoustics, Speech, and Signal Processing (ICASSP 2013)}, Vancouver, Canada, May 2013.
}
\thanks{
The authors gratefully acknowledge the support of the International Graduate School on Mobile Communications 
(MOBICOM), Ilmenau, Germany.
}
\thanks{
The authors J.~Steinwandt, F.~Roemer, M.~Haardt, and G.~Del Galdo are with Ilmenau University of Technology,
P.O.~Box 100565, D-98684 Ilmenau, Germany,
e-mail: \{jens.steinwandt, florian.roemer, martin.haardt, giovanni.delgaldo\}@tu-ilmenau.de,
phone: +49 (3677) 69-2613, web: http://www.tu-ilmenau.de/crl and http://www.tu-ilmenau.de/dvt.
}
\thanks{$*$ corresponding author}
}

\begin{document}

\maketitle
\linespread{1}

\vspace{-1cm}

%
\begin{abstract}
High-resolution parameter estimation algorithms designed to exploit the prior knowledge about 
incident signals from strictly second-order (SO) non-circular (NC) sources allow for a lower estimation error and 
can resolve twice as many sources. In this paper, we derive the $R$-D NC Standard ESPRIT and the $R$-D NC 
Unitary ESPRIT algorithms that provide a significantly better performance compared to their 
original versions for arbitrary source signals. They are applicable to shift-invariant 
$R$-D antenna arrays and do not require a centro-symmetric array structure. Moreover, we present 
a first-order asymptotic performance analysis of the proposed algorithms, which is based 
on the error in the signal subspace estimate arising from the noise perturbation. The 
derived expressions for the resulting parameter estimation error are explicit in the noise 
realizations and asymptotic in the effective signal-to-noise ratio (SNR), i.e., the results become 
exact for either high SNRs or a large sample size. We also provide mean squared error (MSE) 
expressions, where only the assumptions of a zero mean and finite SO moments of the noise are 
required, but no assumptions about its statistics are necessary. As a main result, we analytically 
prove that the asymptotic performance of both $R$-D NC ESPRIT-type algorithms is identical in 
the high effective SNR regime. Finally, a case study shows that no improvement from strictly non-circular 
sources can be achieved in the special case of a single source.
\end{abstract}
%


\IEEEpeerreviewmaketitle


\section{Introduction} 
\label{sec:intro}
\IEEEPARstart{E}{stimating} the parameters of multidimensional ($R$-D) signals with $R\geq1$, e.g., their directions 
of arrival, frequencies, Doppler shifts, etc., has long been of great research interest, given its importance 
in a variety of applications such as radar, sonar, channel sounding, and wireless communications. Among other 
subspace-based parameter estimation schemes (see \cite{krim1996decades,schmidt1986music}), $R$-D Standard ESPRIT 
\cite{roy1989esprit}, $R$-D Unitary ESPRIT \cite{haardt1995unitesprit,haardt1998schur}, and their tensor 
extensions $R$-D Standard Tensor-ESPRIT and $R$-D Unitary Tensor-ESPRIT \cite{haardt2008tensor} are some of the most valuable 
estimators due to their high resolution and their low complexity. However, these methods assume arbitrary 
source signals and do not take prior knowledge such as the second-order (SO) non-circularity of the received 
signals into account. With the growing popularity of subspace-based parameter estimation algorithms, their 
performance analysis has attracted considerable attention. The two most prominent performance 
assessment strategies have been proposed in \cite{rao1989perf} and \cite{vaccaro1993perf}. The concept in 
\cite{rao1989perf} and the follow-up papers \cite{fried1990music,mathews1990music,pillai1989music} analyze
the eigenvector distribution of the sample covariance matrix, originally proposed in \cite{brill1975data}. 
However, it requires Gaussianity assumptions on the source symbols and the noise, and is only asymptotic in 
the sample size $N$. In contrast, \cite{vaccaro1993perf} and its extensions \cite{xu2002perf,liu2008perf} 
provide an explicit first-order approximation of the estimation error caused by the perturbed subspace estimate 
due to a small additive noise contribution. It directly models the leakage of the noise subspace into the signal 
subspace. Unlike \cite{rao1989perf}, this approach is asymptotic in the effective signal-to-noise ratio (SNR), 
i.e., the results become accurate for either high SNRs or a large sample size $N$. Thus, it is even valid for 
the single snapshot case $N=1$ if the SNR is sufficiently high. Furthermore, as it is explicit in the noise 
realizations, no assumptions about the statistics of the signals or the noise are necessary. However, for the 
mean squared error (MSE) expressions in \cite{vaccaro1993perf}, a circularly symmetric noise distribution is 
assumed. In \cite{roemer2012perf} and \cite{RHD:14}, we have derived new MSE expressions that only require the noise to be 
zero-mean with finite SO moments regardless of its statistics and extended the framework of \cite{vaccaro1993perf} 
to the case of $R$-D parameter estimation. Further extensions of these results for the perturbation analyses 
of Tensor-ESPRIT-type algorithms have been presented in \cite{roemer2009hosvd} and \cite{roemer2010perf}, respectively. 
The special case of the performance assessment for a single source was considered in \cite{rao1989perf} and 
the asymptotic efficiency of MUSIC and Root-MUSIC was presented in \cite{porat1988music} and \cite{rao1989root}, 
respectively. However, these results are asymptotic in the sample size $N$ or even in the number of sensors 
$M$. The results presented here are also accurate for small values of $M$ and asymptotic in the effective SNR.

Recently, a number of improved high-resolution subspace-based parameter estimation schemes have been proposed 
for strictly non-circular (NC) sources. These include NC MUSIC \cite{abeida2006perf}, NC Root-MUSIC \cite{charge2001ncroot}, 
1-D NC Standard ESPRIT \cite{zoubir2003ncesprit}, and 2-D NC Unitary ESPRIT \cite{haardt2004ncunit}. Unlike the 
original parameter estimation methods, they exploit prior knowledge about the signals' SO statistics, i.e., their 
strict SO non-circularity \cite{schreier2010noncirc}. Examples of such signals include BPSK, Offset-QPSK, PAM, and 
ASK-modulated signals. By applying a preprocessing procedure similar to the concept of widely-linear processing 
\cite{schreier2010noncirc}, the array aperture is virtually doubled, which results in a significantly reduced 
estimation error and the ability to resolve twice as many sources \cite{haardt2004ncunit}. Some potential applications 
are wireless communications, cognitive radio, etc., when strictly non-circular sources are known to be present, and 
radar, tracking, channel sounding, etc., where the signals can be designed as strictly non-circular signals. The performance 
of NC MUSIC has been derived in \cite{abeida2006perf} based on \cite{rao1989perf}, its source resolvability has been 
investigated in \cite{abeida2008perf}, and mutual coupling has been considered in \cite{huang2010perf}. However, a 
performance analysis of NC Standard ESPRIT and NC Unitary ESPRIT has not yet been reported in the literature. 

In this paper, we first present the $R$-D NC Standard ESPRIT and the $R$-D NC Unitary ESPRIT algorithms as an extension 
of \cite{zoubir2003ncesprit} and \cite{haardt2004ncunit}. They exploit the strict SO non-circularity of stationary 
sources. The algorithms in \cite{zoubir2003ncesprit} and \cite{haardt2004ncunit} are only designed for the case of 1-D 
parameter estimation and require a shift-invariant and centro-symmetric array structure. Here, we relax this requirement 
to only shift-invariant-structured arrays and additionally consider the case of $R$-D ($R\geq1$) parameter estimation. 
Furthermore, we show that the preprocessing step for non-circular sources automatically includes forward-backward 
averaging (FBA) \cite{pillai1989fba}, which is in this case even applicable to arrays without centro-symmetry. In analogy 
to \cite{haardt1995unitesprit}, $R$-D NC Unitary ESPRIT can also be efficiently implemented in terms of only real-valued 
computations by mapping the centro-Hermitian FBA-processed measurement matrix into a real-valued matrix 
\cite{lee1980centro}. This substantially reduces the computational complexity. Regarding the estimation error, $R$-D NC 
Unitary ESPRIT performs better than $R$-D NC Standard ESPRIT at low signal-to-noise ratios (SNR), while simultaneously 
requiring a lower computational load. Both algorithms achieve a significantly lower estimation error than their 
traditional (non-NC) counterparts $R$-D Standard ESPRIT and $R$-D Unitary ESPRIT \cite{haardt1998schur}.

In our second contribution, we extend our initial results in \cite{steinwandt2013ncunit} and derive a first-order 
asymptotic performance analysis of the proposed $R$-D NC Standard and $R$-D NC Unitary ESPRIT algorithms. 
Least squares (LS) is used to solve the resulting augmented shift invariance equations after the preprocessing 
for non-circular sources. Due to its discussed advantages, we resort to the framework in \cite{vaccaro1993perf} 
combined with \cite{RHD:14} for our presented performance analysis. We further extend \cite{RHD:14} 
by incorporating the preprocessing for non-circular sources and derive an explicit first-order expansion of the 
estimation error in terms of the noise perturbation. The noise is assumed to be small compared to the signals but no 
assumptions about its statistics are required. We also provide MSE expressions, where only the assumptions of a 
zero mean and finite SO moments of the noise are needed. Thus, they also give insights into the achievable 
performance in scenarios with non-Gaussian and non-circular perturbations that can, for instance, be caused 
by clutter environments in radar applications \cite{guerci2003stap}. All the obtained expressions are asymptotic 
in the effective SNR, i.e., they become accurate for either high SNRs or large sample sizes. Furthermore, we 
analytically prove that $R$-D NC Standard ESPRIT and $R$-D NC Unitary ESPRIT have the same asymptotic performance 
in the high effective SNR regime. In contrast to \cite{steinwandt2013ncunit}, we here also take the real-valued transformation 
in $R$-D NC Unitary ESPRIT into account for the proof. 

Finally, we present simplified $R$-D MSE expressions for both NC ESPRIT-type algorithms in the special case of a single 
strictly non-circular source, where a uniform sampling grid and circularly symmetric white noise are assumed. The obtained 
closed-form expressions only depend on the physical parameters, i.e., the array size $M$ and the effective SNR. They facilitate design 
decisions on $M$ to achieve a certain performance for specific SNRs. Furthermore, we also simplify the deterministic $R$-D NC 
Cram\'{e}r-Rao bound (CRB) \cite{roemer2007nccrb}\footnote{\!\cite{roemer2007nccrb} only considers the 2-D case, but the 
$R$-D extension is straightforward.} for this case and analytically compute the asymptotic efficiency of the proposed 
algorithms for $R=1$. Note that in \cite{steinwandt2013sls} and \cite{steinwandt2014spasmoo}, we have also incorporated 
structured least squares and spatial smoothing into the performance analysis, respectively.

This remainder of this paper is organized as follows: The data model and the preprocessing for strictly non-circular 
sources are introduced in Section \ref{sec:data} and Section \ref{sec:widely}. In Section \ref{sec:ncunitesprit}, the $R$-D 
NC Standard ESPRIT and $R$-D NC Unitary ESPRIT algorithms are derived. Their performance analysis is presented in Section 
\ref{sec:perf} before the special case of a single source is analyzed in Section \ref{sec:single}. Section 
\ref{sec:simulations} illustrates and discusses the numerical results, and concluding remarks are drawn in Section 
\ref{sec:conclusions}.

\textit{Notation:} We use italic letters for scalars, lower-case bold-face letters for column vectors, 
and upper-case bold-face letters for matrices. The superscripts $^\trans$, $^\conj$, $^\herm$, $^{-1}$, 
and $^+$ denote the transposition, complex conjugation, conjugate transposition, matrix inversion, and the 
Moore-Penrose pseudo inverse of a matrix, respectively. The Kronecker product is denoted as $\otimes$ and 
the Hadamard product is defined as $\odot$. The operator $\vecof{\bm A}$ stacks the columns of the matrix 
$\bm A\in\compl^{M\times N}$ into a column vector of length $MN\times 1$ and $\mathrm{arg}\{\cdot\}$ extracts 
the phase of a complex number. The operator $\mathrm{diag}\{\bm a\}$ returns a diagonal matrix with the 
elements of $\bm a$ placed on its diagonal and $\mathrm{blkdiag}\{\cdot\}$ creates a block diagonal matrix. 
The operator $\mathcal O\{\cdot\}$ denotes the highest order with respect to a parameter. The matrix $\bm 
\Pi_M$ is the $M \times M$ exchange matrix with ones on its antidiagonal and zeros elsewhere. Also, the 
matrices $\bm 1_M$ and $\bm 0_M$ denote the $M \times M$ matrices of ones and zeros, respectively. 
Moreover, $\realof{\cdot}$ and $\imagof{\cdot}$ extract the real and imaginary part of a complex number 
or a matrix respectively, $\twonorm{\bm x}$ represents the 2-norm of the vector $\bm x$, and 
$\expvof{\cdot}$ stands for the statistical expectation. 
\section{Data Model}
\label{sec:data}
Let a noise-corrupted linear superposition of $d$ undamped exponentials be sampled on an arbitrary 
$R$-dimensional ($R$-D) shift-invariant-structured grid\footnote{The grid needs to be decomposable into the 
outer product of $R$ one-dimensional sampling grids \cite{RHD:14}.} of size $M_1\times\ldots\times M_R$ at $N$ 
subsequent time instants \cite{haardt1998schur}. The $t_n$-th time snapshot of the observed $R$-D data 
sequence can be modeled as

\vspace{-1em}
\small
\begin{align}
x_{m_{1},\ldots,m_{R}}(t_n) = \sum_{i=1}^d s_i(t_n) \prod_{r=1}^R \expof{\j k_{m_r}\rd{\mu}_i} + n_{m_{1},\ldots,m_{R}}(t_n),
\label{rdmodel}
\end{align}
\normalsize
where $m_r=1,\ldots,M_r$, $n=1,\ldots,N$, $s_i(t_n)$ denotes the complex amplitude of the 
$i$-th undamped exponential at time instant $t_n$, and $k_{m_r}$ defines the sampling grid\footnote{For a 
uniform sampling grid, we have $k_{m_r}=m_r-1$. An example for a non-uniform grid is provided in 
Fig.~\ref{fig_subsp_2dshinv}, where $k_{m_1}=0,1,2,4,5$ and $k_{m_2}=0,1,3,4$.}. Furthermore, $\rd{\mu}_i$ 
is the spatial frequency in the $r$-th mode for $i=1,\ldots,d$ and $r=1,\ldots,R$, and $n_{m_{1},\ldots,m_{R}}(t_n)$ 
contains the samples of the zero-mean additive noise component. In the array signal processing context, each of the 
$R$-D exponentials represents a narrow-band planar wavefront emitted from stationary far-field 
sources and the complex amplitudes $s_i(t_n)$ are the source symbols. The objective is 
to estimate the $d$ spatial frequencies $\bm \mu_i = [\mu^{(1)}_i,\ldots,\mu^{(R)}_i]^\trans,~\forall i$, 
from \eqref{rdmodel}. We assume that $d$ is known and has been estimated beforehand using model order selection 
techniques, e.g., \cite{wax1985modelorder,huang2007modelorder,huang2013modelorder}. 

In order to obtain a more compact formulation of \eqref{rdmodel}, we collect the observed samples into 
a measurement matrix $\bm X\in\compl^{M\times N}$ with $M=\prod_{r=1}^R M_r$ by stacking the $R$ spatial 
dimensions along the rows and aligning the $N$ time snapshots as the columns. We can then model $\bm X$ as 
\begin{equation}
\bm X=\bm A\bm S+\bm N ~ \in\mathbb C^{M\times N},
\label{model}
\end{equation}
where $\bm A=[\bm a(\bm \mu_1),\ldots,\bm a(\bm \mu_d)]\in\mathbb C^{M\times d}$ is the array steering 
matrix. It consists of the array steering vectors $\bm a(\bm \mu_i)$ corresponding to the $i$-th 
spatial frequency defined by
\begin{equation}
\bm a(\bm \mu_i)= \bm a^{(1)}\left(\mu_i^{(1)}\right)\otimes\cdots\otimes\bm a^{(R)}\left(\mu_i^{(R)}\right)~ \in\mathbb C^{M\times 1},
\label{steer}
\end{equation}
where $\rd{\bm a}\left(\rd{\mu}_i\right)\in\mathbb C^{M_r\times 1}$ is the array steering vector in the $r$-th mode. 
Furthermore, $\bm S\in\mathbb C^{d\times N}$ represents the source symbol matrix and 
$\bm N\in\mathbb C^{M\times N}$ contains the samples of the additive sensor noise. 
Due to the assumption of strictly SO non-circular sources, the complex symbol amplitudes of each source 
form a rotated line in the complex plane so that $\bm S$ can be decomposed as \cite{haardt2004ncunit}
\begin{equation}
\bm S=\bm \Psi\bm S_0,
\label{ncsignal}
\end{equation}
where $\bm S_0\in\mathbb R^{d\times N}$ is a real-valued symbol matrix and $\bm \Psi=
\mathrm{diag}\{[\expof{\j\varphi_1},\ldots,\expof{\j\varphi_d}]^\trans\}\in\mathbb C^{d\times d}$ 
contains stationary complex phase shifts on its diagonal that can be different for 
each source.
\vspace{-0.7em}
\section{Preprocessing for $R$-D NC ESPRIT-Type Algorithms}
\label{sec:widely}
In this section, we derive the NC model resulting from the preprocessing for strictly non-circular sources. We 
show that the shift invariance equations also hold in the NC case and that the virtual 
array always possesses a centro-symmetric structure, even if the physical array is not centro-symmetric. 

In order to take advantage of the benefits associated with strictly non-circular sources, we apply 
a preprocessing procedure and define the augmented measurement matrix $\nc{\bm X}\in\mathbb C^{2M\times N}$ as
\begin{align}
\nc{\bm X}&=\begin{bmatrix} \bm X \\ \bm \Pi_M\bm X^\conj\end{bmatrix}=\begin{bmatrix} \bm A\bm S 
\\ \bm \Pi_M\bm A^\conj\bm S^\conj\end{bmatrix}+\begin{bmatrix} \bm N \\ \bm \Pi_M\bm N^\conj\end{bmatrix}\notag\\ 
&=\begin{bmatrix} \bm A \\ \bm \Pi_M\bm A^\conj\bm \Psi^\conj\bm \Psi^\conj\end{bmatrix}\bm S+
\begin{bmatrix} \bm N \\ \bm \Pi_M\bm N^\conj\end{bmatrix} \label{prop}\\
&= \nc{\bm A} \bm S + \nc{\bm N} = \nc{\bm X_0} + \nc{\bm N},
\label{ncmodel}
\end{align}
where the multiplication by $\bm \Pi_M$ is used to facilitate the real-valued implementation of $R$-D NC Unitary ESPRIT later in 
\eqref{nctrans}. Moreover, $\nc{\bm A}\in\mathbb C^{2M\times d}$ and $\nc{\bm N}\in\mathbb C^{2M\times N}$ 
are the augmented array steering matrix and the augmented noise matrix, respectively, $\nc{\bm X_0}\in\mathbb C^{2M\times N}$ 
is the unperturbed augmented measurement matrix, and we have used the fact that $\bm S_0=\bm \Psi^\conj\bm S$ in 
\eqref{prop}. The extended dimensions of $\nc{\bm A}$ can be interpreted as a virtual doubling of the number of 
sensor elements, which also doubles the number of detectable sources and provides a lower estimation error. 

Based on the assumption that the array steering matrix $\bm A$ is shift-invariant, we next analyze the properties 
of the augmented array steering matrix $\nc{\bm A}$. The shift invariance properties for the physical array 
described by $\bm A$ are given by 
\begin{equation}
\rdtil{\bm J}_1 \bm A ~\rd{\bm \Phi} = \rdtil{\bm J}_2 \bm A, \quad r=1,\ldots,R,
\label{shift}
\end{equation}
where $\rdtil{\bm J}_1$ and $\rdtil{\bm J}_2 \in\real^{\frac{M}{M_r}\subselr{M}{r}\times M}$ are the 
effective $R$-D selection matrices, which select $\subselr{M}{r}$ elements for the first and 
the second subarray in the $r$-th mode, respectively. 
They are compactly defined as $\rdtil{\bm J}_k=\bm I_{\prod_{l=1}^{r-1} M_l} \otimes \rd{\bm J}_k \otimes 
\bm I_{\prod_{l=r+1}^{R} M_l}$ for $k=1,2$, where $\rd{\bm J}_k \in\real^{\subselr{M}{r}\times M_r}$ are 
the $r$-mode selection matrices for the first and second subarray \cite{haardt1998schur}. The diagonal matrix $\rd{\bm \Phi}  =\mathrm{diag}\{[\expof{\j\rd{\mu}_1},\ldots,\expof{\j\rd{\mu}_d}]^\trans\}\in\mathbb C^{d\times d}$ contains 
the spatial frequencies in the $r$-th mode to be estimated. 

The first important property of the augmented steering matrix $\nc{\bm A}$ is formulated in the following 
theorem:
\begin{thm}
If the array steering matrix $\bm A$ is shift-invariant \eqref{shift}, then $\nc{\bm A}$ is also shift-invariant 
and satisfies
\begin{equation}
\ncrtil{\bm J}_1 \nc{\bm A} \rd{\bm \Phi} = \ncrtil{\bm J}_2 \nc{\bm A}, \quad r=1,\ldots,R,
\label{shiftnc}
\end{equation}
where  
\begin{align}
&\ncrtil{\bm J}_k = \bm I_{\prod_{l=1}^{r-1} M_l} \otimes \ncr{\bm J}_k \otimes \bm I_{\prod_{l=r+1}^{R} M_l}, ~k=1,2, \label{j1nctil} \\[0.5ex] 
&\!\!\!\ncr{\bm J}_1 {=} \mathrm{blkdiag}\left\{ \rd{\bm J}_1, \bm \Pi_{M_r^{(\mathrm{sel})}} \rd{\bm J}_2 \bm \Pi_{M_r}\right\} 
\in\mathbb R^{2M_r^{(\mathrm{sel})}\times 2M_r}, \notag\\
&\!\!\!\ncr{\bm J}_2 {=} \mathrm{blkdiag}\left\{ \rd{\bm J}_2, \bm \Pi_{M_r^{(\mathrm{sel})}} \rd{\bm J}_1 \bm \Pi_{M_r}\right\} 
\in\mathbb R^{2M_r^{(\mathrm{sel})}\times 2M_r}. \notag 
\end{align}
\label{thm:shift}
\end{thm} 
\vspace{-1.5em}
\begin{IEEEproof}
See Appendix \ref{app:shift}.
\end{IEEEproof}
If the physical array is centro-symmetric, i.e., it is symmetric with respect to its 
centroid, its array steering matrix $\bm A_\mathrm{c}$ satisfies \cite{haardt1995unitesprit}
\begin{equation}
\bm \Pi^{}_M\bm A_\mathrm{c}^\conj=\bm A^{}_\mathrm{c}\bm \Delta_\mathrm{c},
\label{centro}
\end{equation}
where $\bm \Delta_\mathrm{c}\in\mathbb C^{d\times d}$ is a unitary diagonal matrix\footnote{In case of a 
physical centro-symmetric array, $\bm \Delta_\mathrm{c}$ depends on the phase center of the array. If the 
phase center coincides with the array's centroid, we have $\bm \Delta_\mathrm{c}=\bm I_d$.}. If \eqref{centro} 
holds, we have $\rd{\bm J}_2 = \bm \Pi_{\subsel{M_r}} \rd{\bm J}_1 \bm \Pi_{M_r}$ and hence the augmented 
selection matrices $\ncr{\bm J}_1$ and $\ncr{\bm J}_2$ simplify to 
\begin{equation}
\ncr{\bm J}_k=\bm I^{}_2 \otimes \rd{\bm J}_k,~k=1,2.
\label{jncula}
\end{equation}
Note that this special case was assumed in \cite{zoubir2003ncesprit} and \cite{haardt2004ncunit}.

The second important property of $\nc{\bm A}$ is stated in the following theorem:
\begin{thm}
The augmented steering matrix $\nc{\bm A}$ always exhibits centro-symmetry 
even if $\bm A$ is not centro-symmetric.
\label{thm:centro}
\end{thm}
\vspace{-1.5em}
\begin{IEEEproof}
Assuming that $\bm A$ does not necessarily satisfy \eqref{centro}, we have
\begin{align}
\bm \Pi_{2M} \ncconj{\bm A} &= \begin{bmatrix} \bm 0 & \bm \Pi_M \\ \bm \Pi_M &\bm 0\end{bmatrix}
\begin{bmatrix} \bm A^\conj  \\ \bm \Pi_M\bm A\bm \Psi\bm \Psi\end{bmatrix} 
=\begin{bmatrix} \bm A\bm \Psi\bm \Psi \\ \bm \Pi_M\bm A^\conj \end{bmatrix} \notag\\
&=\begin{bmatrix} \bm A \\ \bm \Pi_M\bm A^\conj\bm \Psi^\conj\bm \Psi^\conj \end{bmatrix}\bm \Psi\bm \Psi 
= \nc{\bm A} \bm \Delta_\mathrm{c},
\label{centronc}
\end{align}
where $\bm \Delta_\mathrm{c}$ becomes $\bm \Psi\bm \Psi$, which is unitary and diagonal. Therefore, 
$\nc{\bm A}$ satisfies \eqref{centro}, which shows that it is centro-symmetric regardless of the 
centro-symmetry of $\bm A$.
\end{IEEEproof}

This result shows that $R$-D NC Unitary ESPRIT, derived in the next section, can be applied to a 
broader variety of array geometries than $R$-D Unitary ESPRIT, which requires a centro-symmetric array. 
An example is provided in Fig.~\ref{fig:rmse_snap} of Section \ref{sec:simulations}.
\section{Proposed $R$-D NC ESPRIT-Type Algorithms}
\label{sec:ncunitesprit}
In this section, we present the NC Standard ESPRIT and the NC Unitary ESPRIT algorithms for arbitrarily 
formed $R$-dimensional shift-invariant-structured array geometries, where centro-symmetry is not required. 
Furthermore, we summarize some important properties at the end.
\subsection{$R$-D NC Standard ESPRIT Algorithm}
Based on the noisy augmented data model \eqref{ncmodel}, we estimate the signal subspace $\hat{\bm U}_
\mathrm{s}^{(\mathrm{nc})}\in\mathbb C^{2M\times d}$ by computing the $d$ dominant left singular vectors 
of $\bm X^{(\mathrm{nc})}$. As $\bm A^{(\mathrm{nc})}$ and $\hat{\bm U}_\mathrm{s}^{(\mathrm{nc})}$ span 
approximately the same column space, we can find a non-singular matrix $\bm T\in\mathbb C^{d\times d}$ 
such that $\bm A^{(\mathrm{nc})}\approx\hat{\bm U}_\mathrm{s}^{(\mathrm{nc})}\bm T$. Using this relation, 
the overdetermined set of $R$ augmented shift invariance equations \eqref{shiftnc} can be expressed in 
terms of the estimated augmented signal subspace, yielding
\begin{equation}
\ncrtil{\bm J}_1 \nc{\hat{\bm U}}_\mathrm{s} \rd{\bm \Gamma} \approx \ncrtil{\bm J}_2 \nc{\hat{\bm U}}_\mathrm{s}, \quad r=1,\ldots,R
\label{shiftncsubstan}
\end{equation}
with $\rd{\bm \Gamma}=\bm T\rd{\bm \Phi}\bm T^{-1}$. Often, the $R$ unknown matrices $\rd{\bm \Gamma}\in\compl^{d\times d}$ 
are estimated using least squares (LS), i.e.,
\begin{equation}
\rd{\hat{\bm \Gamma}}=\left(\ncrtil{\bm J}_1 \nc{\hat{\bm U}}_\mathrm{s}\right)^+ \ncrtil{\bm J}_2 \nc{\hat{\bm U}}_\mathrm{s} 
\in\compl^{d\times d}.
\label{lsstan}
\end{equation}
Finally, after solving \eqref{lsstan} for $\rd{\hat{\bm \Gamma}}$ in each mode independently, the correctly paired 
spatial frequency estimates are given by $\rd{\hat{\mu}}_i=\mathrm{arg}\{\rd{\hat{\lambda}}_i\},~i=1,\ldots,d$. 
The eigenvalues $\rd{\hat{\lambda}}_i$ of $\rd{\hat{\bm \Gamma}}$ are obtained by performing a joint 
eigendecomposition across all $R$ dimensions \cite{fu2006diag} or via the simultaneous Schur decomposition 
\cite{haardt1998schur}. The $R$-D NC Standard ESPRIT algorithm is summarized in Table \ref{tab:ncesprit}.
\begin{table}[!t]
\centering
\caption{Summary of the $R$-D NC Standard ESPRIT Algorithm}     
\label{tab:ncesprit}
\begin{small} 
\vspace{-2mm}
\begin{tabular}{|p{8.1cm}|} 
\hline \\[-2ex]
\begin{enumerate} \vspace{-2mm}
	\item Estimate the augmented signal subspace $\nc{\hat{\bm U}}_s \in\compl^{2M\times d}$ via the truncated 
	SVD of the augmented observation $\nc{\bm X}\in\compl^{2M\times N}$. \vspace{1mm}
	\item Solve the overdetermined set of augmented shift invariance equations 
	\begin{align} \notag
	\ncrtil{\bm J}_1 \nc{\hat{\bm U}}_\mathrm{s} \rd{\bm \Gamma} \approx \ncrtil{\bm J}_2 \nc{\hat{\bm U}}_\mathrm{s}
	\end{align}
	for $\rd{\bm \Gamma}\in\compl^{d\times d},~r=1,\ldots,R,$
	by using an LS algorithm, where $\ncrtil{\bm J}_k \in\real^{\frac{M}{M_r}\subselr{M}{r}\times 2M},~k=1,2$, is defined in \eqref{j1nctil}. 
	\item Compute the eigenvalues $\rd{\hat{\lambda}}_i,~i=1,\ldots,d$ of $\rd{\bm \Gamma}$ jointly for all $r=1,\ldots,R$. 
	Recover the correctly paired spatial frequencies $\rd{\hat{\mu}}_i$ via 
	\begin{align} \notag
	\rd{\hat{\mu}}_i = \mathrm{arg}\{\rd{\hat{\lambda}}_i\}.
	\end{align} 
  \vspace{-2.5em} \end{enumerate} \\
\hline 
\end{tabular}
\end{small} 
\end{table}
\subsection{$R$-D NC Unitary ESPRIT Algorithm}
As a main feature, $R$-D Unitary ESPRIT involves forward-backward averaging (FBA) \cite{pillai1989fba} 
of the measurement matrix $\bm X$, which results in a centro-Hermitian matrix, i.e, matrices $\bm Z\in
\mathbb C^{p\times q}$ that satisfy $\bm \Pi_p\bm Z^\conj \bm \Pi_q = \bm Z$. Therefore, it can be 
efficiently formulated in terms of only real-valued computations \cite{haardt1995unitesprit}. This is 
achieved by a bijective mapping of the set of centro-Hermitian matrices onto the set of real-valued 
matrices \cite{lee1980centro}. To this end, let us define left $\bm \Pi$-real matrices, i.e., matrices 
$\bm Q\in\mathbb C^{p\times q}$ satisfying $\bm \Pi_p\bm Q^\conj=\bm Q$. A sparse and square unitary 
left $\bm \Pi$-real matrix of odd order is given by
\begin{equation}
\bm Q_{2n+1}=\frac{1}{\sqrt{2}} \cdot
\begin{bmatrix}
\bm I_n & \bm 0_{n\times 1} & \j\bm I_n \\
\bm 0_{n\times 1}^\trans & \sqrt{2} & \bm 0_{n\times 1}^\trans \\
\bm \Pi_n & \bm 0_{n\times 1} & -\j\bm \Pi_n \\
\end{bmatrix}.
\label{leftq}
\end{equation}
A unitary left $\bm \Pi$-real matrix of even order is obtained from \eqref{leftq} by dropping its 
center row and center column. More left $\bm \Pi$-real matrices can be constructed by post-multiplying 
a left $\bm \Pi$-real matrix $\bm Q$ by an arbitrary real matrix $\bm R$ of appropriate size. Using 
this definition, any centro-Hermitian matrix $\bm Z\in\mathbb C^{p\times q}$ can be transformed into 
a real-valued matrix through the transformation \cite{lee1980centro}
\begin{equation}
\varphi(\bm Z) = \bm Q^\herm_p \bm Z \bm Q^{}_q \in\real^{p\times q}.
\label{realval}
\end{equation}
In Unitary ESPRIT, the centro-Hermitian matrix obtained after FBA is given by 
\cite{haardt1995unitesprit}
\begin{equation}
\tilde{\bm X}=\begin{bmatrix} \bm X & \bm \Pi_{M}\bm X^\conj\bm \Pi_N \end{bmatrix} \in\compl^{M\times 2N}.
\label{fba}
\end{equation}

Next, we extend the concept of Unitary ESPRIT to the augmented data model in \eqref{ncmodel} and derive the $R$-D
NC Unitary ESPRIT algorithm. Therefore, the FBA step as well as the real-valued transformation have to be applied 
to $\nc{\bm X}$. Here, FBA is performed by replacing the NC measurement matrix $\nc{\bm X}\in\mathbb C^{2M\times N}$ 
by the column-wise augmented measurement matrix $\ncfba{\bm X}\in\mathbb 
C^{2M\times 2N}$ defined by
\begin{align}
\ncfba{\bm X}&=\begin{bmatrix} \nc{\bm X} & \bm \Pi_{2M}\ncconj{\bm X}\bm \Pi_N \end{bmatrix} \label{fbaa}\\
&=\begin{bmatrix} \bm X & \bm X\bm \Pi_N \\ \bm \Pi_{M}\bm X^\conj & \bm \Pi_{M}\bm X^\conj\bm \Pi_N \end{bmatrix} \notag\\
&=\begin{bmatrix} \nc{\bm X} & \nc{\bm X}\bm \Pi_N  \end{bmatrix}.
\label{fbanc}
\end{align}
Due to the fact that equivalently to \eqref{fba}, $\ncfba{\bm X}$ is centro-Hermitian, it can be 
transformed into a real-valued matrix that takes the simple form 
\begin{align}
\varphi(\ncfba{\bm X})&=\bm Q^\herm_{2M}\ncfba{\bm X}\bm Q_{2N} \label{nctrans1}\\
&=2\cdot\begin{bmatrix} \realof{\bm X} & \bm 0_{M\times N}\\ \imagof{\bm X} & \bm 0_{M\times N} \end{bmatrix}.
\label{nctrans}
\end{align}
The proof is given in Appendix \ref{app:trans}. 

In the next step, we define the transformed augmented steering matrix as $\nc{\bm D} = \bm Q^\herm_{2M} \nc{\bm A}$. 
Based on the $R$-D shift invariance property of $\nc{\bm A}$ proven in Theorem \ref{thm:shift}, it can easily be 
verified that $\nc{\bm D}$ obeys
\begin{align}
\ncrtil{\bm K}_1 \nc{\bm D} \rd{\bm \Omega} = \ncrtil{\bm K}_2 \nc{\bm D}, \quad r=1,\ldots,R,
\label{shiftncreal}
\end{align}
where the $R$ pairs of augmented selection matrices in \eqref{j1nctil} are transformed according to \cite{haardt1995unitesprit} 
as
\begin{align}
\ncrtil{\bm K}_1 & = 2\cdot\realof{\bm Q^\herm_{\frac{M}{M_r}\subselr{M}{r}} \ncrtil{\bm J}_2 \bm Q^{}_{2M}} \label{realsel1} \\
\ncrtil{\bm K}_2 & = 2\cdot\imagof{\bm Q^\herm_{\frac{M}{M_r}\subselr{M}{r}} \ncrtil{\bm J}_2 \bm Q^{}_{2M}}.
\label{realsel2}
\end{align}
Moreover, the real-valued set of diagonal matrices $\rd{\bm \Omega} = \mathrm{diag}\{[\rd{\omega}_1,\ldots,\rd{\omega}_d]^\trans\} 
\in\real^{d\times d}$ with $\rd{\omega}_i=\tan(\rd{\mu}_i/2)$ contain the spatial frequencies in the $r$-th mode.

Using the preprocessed noisy data in \eqref{nctrans}, we then estimate the real-valued augmented signal subspace $\nc{\hat{\bm E}}
_{\rm s} \in\real^{2M\times d}$ by computing the $d$ dominant left singular vectors of $\varphi(\ncfba{\bm X})$. 
Note that the zero block matrices and the scaling factor of 2 in \eqref{nctrans} can be dropped as they do not 
alter the signal subspace of $\varphi(\ncfba{\bm X})$. As $\nc{\bm D}$ and $\nc{\hat{\bm E}}_{\rm s}$ span 
approximately the same column space, we can find a non-singular matrix $\bm T\in\compl^{d\times d}$ such that 
$\nc{\bm D} \approx \nc{\hat{\bm E}}_{\rm s} \bm T$. Substituting this relation into \eqref{shiftncreal}, the 
overdetermined set of $R$ real-valued shift invariance equations in terms of the estimated augmented signal 
subspace is given by
\begin{equation}
\ncrtil{\bm K}_1 \nc{\hat{\bm E}_\mathrm{s}} \rd{\bm \Upsilon} \approx \ncrtil{\bm K}_2 \nc{\hat{\bm E}_\mathrm{s}}, \quad r=1,\ldots,R
\label{shiftncsub}
\end{equation}
with $\rd{\bm \Upsilon}=\bm T \rd{\bm \Omega} \bm T^\inv$. Often, the $R$ unknown real-valued diagonal matrices $\rd{\bm \Upsilon}$ 
are estimated using least squares (LS), i.e.,
\begin{equation}
\rd{\hat{\bm \Upsilon}} = \left(\ncrtil{\bm K}_1 \nc{\hat{\bm E}_\mathrm{s}}\right)^+\ncrtil{\bm K}_2 
\nc{\hat{\bm E}_\mathrm{s}}\in\real^{d\times d}.
\label{ls}
\end{equation}
Finally, the correctly paired spatial frequency estimates 
are obtained by $\rd{\hat{\mu}}_i = 2 \cdot \mathrm{arctan}(\rd{\hat{\omega}}_i),~i=1,\ldots,d$. The eigenvalues 
$\rd{\hat{\omega}}_i$ of $\rd{\hat{\bm \Upsilon}}$ are computed by performing a joint 
eigendecomposition across all $R$ dimensions \cite{fu2006diag} or via the simultaneous Schur decomposition 
\cite{haardt1998schur}. If all the eigenvalues are real, they provide reliable 
estimates \cite{haardt1995unitesprit}. A summary of $R$-D NC Unitary ESPRIT is given in Table \ref{tab:ncunitesprit}.
\begin{table}[!t]
\centering
\caption{Summary of the $R$-D NC Unitary ESPRIT Algorithm}     
\label{tab:ncunitesprit}
\begin{small} 
\vspace{-2mm}
\begin{tabular}{|p{8.1cm}|} 
\hline \\[-2ex]
\begin{enumerate} \vspace{-2mm}
	\item Estimate the augmented real-valued signal subspace $\nc{\hat{\bm E}}_s \in\mathbb{R}^{2M\times d}$ via the truncated 
	SVD of the stacked observation 
	\begin{align} \notag
	[\realof{\bm X}^\trans, \imagof{\bm X}^\trans]^\trans \in\mathbb{R}^{2M\times N}. 
	\end{align} 
	\item Solve the overdetermined set of augmented shift invariance equations 
	\begin{align} \notag
	\ncrtil{\bm K}_1 \nc{\hat{\bm E}_\mathrm{s}} 
	\rd{\bm \Upsilon} \approx \ncrtil{\bm K}_2 \nc{\hat{\bm E}_\mathrm{s}}
	\end{align} for $\rd{\bm \Upsilon}\in\real^{d\times d},~r=1,\ldots,R,$
	by using an LS algorithm, where $\ncrtil{\bm K}_k \in\real^{\frac{M}{M_r}\subselr{M}{r}\times 2M},~k=1,2$ and $\ncrtil{\bm J}_2$ are defined 
	in \eqref{realsel1}, \eqref{realsel2}, and \eqref{j1nctil}, respectively. \vspace{1mm}
	\item Compute the eigenvalues $\rd{\hat{\omega}}_i,~i=1,\ldots,d$ of $\rd{\bm \Upsilon}$ jointly for all $r=1,\ldots,R$. 
	Recover the correctly paired spatial frequencies $\rd{\hat{\mu}}_i$ via 
	\begin{align} \notag
	\rd{\hat{\mu}}_i = 2 \cdot \mathrm{arctan}(\rd{\hat{\omega}}_i).
  \end{align}
  \vspace{-2.5em} \end{enumerate} \\
\hline 
\end{tabular}
\end{small} 
\vspace{-1em}
\end{table}
\subsection{Properties of $R$-D NC ESPRIT-Type Algorithms}
The proposed $R$-D NC Standard ESPRIT and $R$-D NC Unitary ESPRIT algorithms have a number of important properties that 
are summarized in this subsection. Firstly, both algorithms can be applied to estimate the parameters 
of stationary strictly SO non-circular sources via shift-invariant $R$-D arrays, where a centro-symmetric array 
structure is not required as shown in Theorem \ref{thm:shift} and Theorem \ref{thm:centro}. Secondly, it will 
be shown in Section \ref{sec:ncunit} that the performance of $R$-D NC Standard ESPRIT and $R$-D NC Unitary ESPRIT 
is asymptotically identical. This is due to the fact that for $R$-D NC Unitary ESPRIT, applying FBA to 
$\nc{\bm X}$ does not improve the signal subspace estimate and the real-valued transformation has no 
effect on the asymptotic performance. As a consequence, $R$-D NC Unitary ESPRIT cannot handle coherent 
sources as FBA has no decorrelation effect. However, spatial smoothing \cite{haardt2004ncunit} can be 
applied to separate coherent wavefronts. Therefore, and thirdly, $R$-D NC Standard ESPRIT and $R$-D NC 
Unitary ESPRIT can both resolve up to 
\begin{align}
\min\big\{\min_{r} (2\cdot\subselr{M}{r}M/M_r),N\big\} 
\end{align}
incoherent sources as compared to $\min\{\min_{r} (\subselr{M}{r}M/M_r),$ $N\}$ and $\min\{\min_{r} 
(\subselr{M}{r}M/M_r),2\cdot N\}$ for $R$-D Standard ESPRIT and $R$-D Unitary ESPRIT, respectively. Thus, if 
$N$ is large enough, we can detect twice as many incoherent sources. Fourth, due to the exchange matrix 
$\bm \Pi_M$ in \eqref{ncmodel}, the real-valued transformation in $R$-D NC Unitary ESPRIT can be efficiently 
computed by stacking the real part and the imaginary part of $\bm X$ on top of each other, cf. equation 
\eqref{nctrans}. Finally, the computational complexity of both algorithms is dominated by the signal subspace 
estimate via the SVD of \eqref{nctrans}, which is of cost $\mathcal O((2M)^2N)$ \cite{golub1996matrix}, and 
the pseudo inverse in \eqref{lsstan} and \eqref{ls}, whose computational cost is $\mathcal O((2M)^3)$ 
\cite{golub1996matrix}. However, the complexity of $R$-D NC Unitary ESPRIT is lower than that of $R$-D NC 
Standard ESPRIT as these operations are real-valued.
\section{Performance of $R$-D NC ESPRIT-Type Algorithms}
\label{sec:perf}
In this section, we present the first-order analytical performance assessment of $R$-D NC Standard ESPRIT and $R$-D 
NC Unitary ESPRIT. As will be shown in Subsection \ref{sec:ncunit}, the performance of $R$-D NC Standard ESPRIT and 
$R$-D NC Unitary ESPRIT is asymptotically identical. Therefore, we first resort to the simpler derivation of 
the expressions for $R$-D NC Standard ESPRIT and then show their equivalence. In contrast to our previous results 
in \cite{steinwandt2013ncunit}, we here also include the real-valued transformation in $R$-D NC Unitary ESPRIT 
into the proof.
\subsection{Performance of $R$-D NC Standard ESPRIT}
\label{sec:SEperf}
To obtain a first-order perturbation analysis of the parameter estimates, we adopt the analytical performance framework 
proposed in \cite{vaccaro1993perf}. Thus, we first develop a first-order subspace error expansion 
in terms of the perturbation $\nc{\bm N}$ and then find a corresponding first-order expansion for the parameter 
estimation error $\Delta\mu_i$. It is evident from \eqref{ncmodel} that the preprocessing does not violate the assumption of a 
small noise perturbation made in \cite{vaccaro1993perf}. Hence, we can apply the concept of \cite{vaccaro1993perf} to 
the augmented measurement matrix in \eqref{ncmodel}. The results are asymptotic in the high effective SNR and 
explicit in the noise term $\nc{\bm N}$. 

Starting with the subspace error expression based on \eqref{ncmodel}, we express the SVD 
of the noise-free observations $\nc{\bm X_0}$ as
\begin{align}
\nc{\bm X_0} = \begin{bmatrix} \nc{\bm U_\mathrm{s}} & \nc{\bm U_\mathrm{n}}\end{bmatrix}
\begin{bmatrix} \nc{\bm \Sigma_\mathrm{s}} & \bm 0 \\ \bm 0 & \bm 0\end{bmatrix}
\begin{bmatrix} \nc{\bm V_\mathrm{s}} & \nc{\bm V_\mathrm{n}}\end{bmatrix}^\herm,
\notag
\end{align}
where $\nc{\bm U_\mathrm{s}}\in\mathbb C^{2M\times d}$, $\nc{\bm U_\mathrm{n}}\in\mathbb C^{2M\times (2M-d)}$, 
and $\nc{\bm V_\mathrm{s}}\in\mathbb C^{N\times d}$ span the signal subspace, the noise subspace, and the 
row space respectively, and $\nc{\bm \Sigma_\mathrm{s}}\in\mathbb R^{d\times d}$ contains the non-zero singular 
values on its diagonal. Next, we write the perturbed signal subspace estimate of $\nc{\hat{\bm U}_\mathrm{s}}$ 
from the previous section as $\nc{\hat{\bm U}_\mathrm{s}}=\nc{\bm U_\mathrm{s}} + \Delta \nc{\bm U_\mathrm{s}}$, 
where $\Delta \nc{\bm U_\mathrm{s}}$ denotes the estimation error. From \cite{vaccaro1993perf} and its 
application to \eqref{ncmodel}, we obtain the first-order subspace error approximation
\begin{equation}
\Delta \nc{\bm U_\mathrm{s}} = \nc{\bm U_\mathrm{n}} \ncH{\bm U_\mathrm{n}}
\nc{\bm N} \nc{\bm V_\mathrm{s}} \ncinv{\bm \Sigma_\mathrm{s}} + \mathcal O\{\nu^2\},
\label{subpert}
\end{equation}
where $\nu=\|\nc{\bm N}\|$, and $\|\cdot\|$ represents an arbitrary sub-multiplicative\footnote{A 
matrix norm is called sub-multiplicative if $\|\bm A\cdot \bm B \|\leq \|\bm A\|\cdot \|\bm B\|$ for 
arbitrary matrices $\bm A$ and $\bm B$.}~norm. Equation \eqref{subpert} models the leakage of the noise 
subspace into the signal subspace due to the effect of the noise. The perturbation of the particular 
basis for the signal subspace $\nc{\bm U_\mathrm{s}}$, which is taken into account in \cite{xu2002perf}, 
\cite{liu2008perf} can be ignored as the choice of this basis is irrelevant for $R$-D NC Standard ESPRIT. 

For the parameter estimation error of the $i$-th spatial frequency in the $r$-th mode obtained by the 
LS solution in \eqref{lsstan}, we follow the lines of \cite{vaccaro1993perf} to obtain
\begin{eqnarray}
\begin{aligned}
\rd{\Delta\mu_i} &=\mathrm{Im}\left\{\bm p_i^\trans \left(\ncrtil{\bm J}_1 \nc{\bm U}_\mathrm{s}\right)^+
\left[\ncrtil{\bm J}_2/\rd{\lambda}_i\right.\right.\\
&\qquad~~\qquad \left.\left.-\ncrtil{\bm J}_1\right]\Delta \nc{\bm U_\mathrm{s}}\bm q_i\right\}+ \mathcal O\{\nu^2\},
\label{estpert}
\end{aligned}
\end{eqnarray}
where $\rd{\lambda}_i=\expof{\j\mu_i}$ is the $i$-th eigenvalue of $\rd{\bm \Gamma}$ in the $r$-th mode, 
$\bm q_i$ represents the $i$-th eigenvector of $\rd{\bm \Gamma}$, i.e., the $i$-th column vector of the 
eigenvector matrix $\bm Q$, and $\bm p_i^\trans$ is the $i$-th row vector of $\bm P=\bm Q^{-1}$. Hence, 
the eigendecomposition of $\rd{\bm \Gamma}$ in the $r$-th mode is given by
\begin{equation}
\rd{\bm \Gamma} = \bm Q \rd{\bm \Lambda} \bm Q^{-1},
\label{psi}
\end{equation}
where $\rd{\bm \Lambda}$ contains the eigenvalues $\rd{\lambda}_i$ on its diagonal. Then, by inserting 
\eqref{subpert} into \eqref{estpert}, we can write the first-order approximation for the estimation errors 
$\rd{\Delta \mu_i}$ explicitly in terms of the noise perturbation $\nc{\bm N}$.

In order to derive an analytical expression for the MSE of $R$-D NC Standard ESPRIT, we resort to 
\cite{RHD:14}, where we have derived an MSE expression that only depends on the 
SO statistics of the noise, i.e., the covariance matrix and the pseudo-covariance matrix, assuming 
the noise to be zero-mean. 
As the preprocessing in \eqref{ncmodel} does not violate the zero-mean assumption, \cite{RHD:14} 
is applicable once the corresponding SO statistics are found. Therefore, defining $\nc{\bm n}=\mathrm{vec}\{\nc{\bm N}\}
\in\mathbb C^{2MN\times 1}$, its covariance matrix $\nc{\bm R_\mathrm{nn}}=\mathbb E\{
\nc{\bm n}\ncH{\bm n}\}\in\mathbb C^{2MN\times 2MN}$, and its pseudo-covariance matrix $\nc{\bm C_\mathrm{nn}}
=\mathbb E\{\nc{\bm n}\ncT{\bm n}\}\in\mathbb C^{2MN\times 2MN}$, the MSE for the $i$-th spatial frequency 
in the $r$-th mode is given by

\small
\vspace{-1em}
\begin{eqnarray}
\begin{aligned}
&\mathbb E\left\{(\rd{\Delta \mu_i})^2\right\} = \frac{1}{2} \left(\ncrH{\bm r}_i \ncconj{\bm W} \ncT{\bm R_\mathrm{nn}} 
\ncT{\bm W} \ncr{\bm r}_i\right.\\
&\left. -\mathrm{Re}\left\{\ncrT{\bm r}_i \nc{\bm W} \ncT{\bm C_\mathrm{nn}} \ncT{\bm W} \ncr{\bm r}_i \right\}\right)
+\mathcal O\{\nu^2\},
\label{mse}
\end{aligned}
\end{eqnarray}
\normalsize
where
\small
\begin{align} \notag
\ncr{\bm r}_i &= \bm q_i\otimes\Big(\Big[\left(\ncrtil{\bm J}_1 \nc{\bm U_\mathrm{s}}\right)^+ \Big.\\
&\quad\Big. \cdot\left(\ncrtil{\bm J}_2/\rd{\lambda}_i - \ncrtil{\bm J}_1\right)\Big]^\trans\bm p_i\Big) ~\in\compl^{2Md\times 1} \notag
\end{align}
\normalsize
and
\small
\begin{equation}\notag
\nc{\bm W}=\left(\ncinv{\bm \Sigma_\mathrm{s}} \ncT{\bm V_\mathrm{s}}\right) \otimes 
\left(\nc{\bm U_\mathrm{n}} \ncH{\bm U_\mathrm{n}}\right) ~\in\compl^{2Md\times 2MN}.
\end{equation}
\normalsize

In the next step, we derive the covariance matrix and the pseudo-covariance matrix of the augmented noise 
contribution $\nc{\bm n}$ required in \eqref{mse}. To this end, we use the commutation 
matrix $\bm K_{M,N}$ of size $MN\times MN$, which is defined as the unique permutation matrix satisfying \cite{magnus1995matrix}
\begin{equation}
\bm K_{M,N}\cdot\mathrm{vec}\{\bm A\}=\mathrm{vec}\{\bm A^\trans\}
\label{comm}
\end{equation}
for arbitrary matrices $\bm A\in\mathbb C^{M\times N}$. We first expand $\nc{\bm n}$ as 
\begin{align}
\nc{\bm n} &=\mathrm{vec}\{\nc{\bm N}\}=\mathrm{vec}\left\{\begin{bmatrix} \bm N \\ \bm \Pi_M\bm N^\conj\end{bmatrix}\right\} \label{noise1}\\
&=\bm K_{2M,N}^\trans\begin{bmatrix} \mathrm{vec}\{\bm N^\trans\} \\ \mathrm{vec}\{(\bm \Pi_M\bm N^\conj)^\trans\} \end{bmatrix}\label{noise2}\\
&=\bm K_{2M,N}^\trans\begin{bmatrix} \bm K_{M,N}\cdot\mathrm{vec}\{\bm N\} \\ \bm K_{M,N}\cdot\mathrm{vec}\{\bm \Pi_M\bm N^\conj\} \end{bmatrix} \notag\\
&=\bm K_{2M,N}^\trans\left(\bm I_2\otimes\bm K_{M,N} \right)\begin{bmatrix} \mathrm{vec}\{\bm N\} \\ \mathrm{vec}\{\bm \Pi_M\bm N^\conj\} \end{bmatrix},
\label{noise}
\end{align}
where we have applied property \eqref{comm} to the equations 
\eqref{noise1} and \eqref{noise2}. By defining $\bm n=\mathrm{vec}\{\bm N\}\in\mathbb C^{MN\times 1}$ 
and using the property $\mathrm{vec}\{\bm A\bm X\bm B\}=(\bm B^\trans\otimes\bm A)\cdot\mathrm{vec}\{\bm X\}$ 
for arbitrary matrices $\bm A$, $\bm B$, and $\bm X$ of appropriate sizes, we can formulate \eqref{noise} 
as 
\begin{equation}
\nc{\bm n}=\tilde{\bm K}\begin{bmatrix} \bm n \\ \bm n^\conj \end{bmatrix},
\label{noisenc}
\end{equation}
where $\tilde{\bm K}=\bm K_{2M,N}^\trans \cdot \mathrm{blkdiag}\{\bm K_{M,N}$,\,$\bm K_{M,N}
\left(\bm I_{N} \otimes \bm \Pi_{M} \right)\}$ is of size $2MN\times 2MN$.
Thus, the SO statistics of $\nc{\bm n}$ can be expressed by means of the covariance matrix 
$\bm R_\mathrm{nn}=\mathbb E\{\bm n\bm n^\herm\}$ and the pseudo-covariance matrix $\bm C_\mathrm{nn}=\mathbb E
\{\bm n\bm n^\trans\}$ of the physical noise $\bm n$. Therefore, we obtain 
\begin{align}
\nc{\bm R_\mathrm{nn}}=\mathbb E\left\{\nc{\bm n} \ncH{\bm n}\right\}=\tilde{\bm K}\begin{bmatrix} \bm R_\mathrm{nn} & \bm C_\mathrm{nn} \\ 
\bm C_\mathrm{nn}^\conj & \bm R_\mathrm{nn}^\conj \end{bmatrix}\tilde{\bm K}^\herm
\label{cov_nc}
\end{align}
and 
\begin{align}
\nc{\bm C_\mathrm{nn}}=\mathbb E\left\{\nc{\bm n} \ncT{\bm n}\right\}=\tilde{\bm K}\begin{bmatrix} \bm C_\mathrm{nn} & \bm R_\mathrm{nn} \\ 
\bm R_\mathrm{nn}^\conj & \bm C_\mathrm{nn}^\conj \end{bmatrix}\tilde{\bm K}^\trans.
\label{pcov_nc}
\end{align} 
In the special case of circularly symmetric white noise with $\bm R_\mathrm{nn}=\sigma^2_\mathrm{n}
\bm I^{}_{MN}$ and $\bm C_\mathrm{nn}=\bm 0_{MN}$, \eqref{cov_nc} and \eqref{pcov_nc} simplify to 
\begin{equation}
\nc{\bm R_\mathrm{nn}}=\sigma^2_\mathrm{n}\bm I^{}_{2MN} \quad\mathrm{and}\quad 
\nc{\bm C_\mathrm{nn}}=\sigma^2_\mathrm{n}(\bm I_N\otimes\bm \Pi_{2M}).
\label{wcsnoise}
\end{equation}
Note that the pseudo-covariance matrix $\nc{\bm C}_\mathrm{nn}$ is always non-zero even in the case of 
circularly symmetric white noise. This is due to the preprocessing in \eqref{ncmodel}.
Furthermore, it is worth mentioning that the step of solving the $R$ augmented shift invariance equations 
for $\rd{\bm \Gamma}$ independently and then performing a joint eigendecomposition across all $R$ dimensions 
to obtain $\rd{\bm \Lambda}$ has no impact on the asymptotic estimation error for high SNRs since the 
eigenvectors become asymptotically equal \cite{RHD:14}.
\subsection{Performance of $R$-D NC Unitary ESPRIT}
\label{sec:ncunit}
So far, we have only derived the explicit first-order parameter estimation error approximation and 
the MSE expression for $R$-D NC Standard ESPRIT. In this subsection, however, we show that the 
analytical performance of $R$-D NC Unitary ESPRIT and $R$-D NC Standard ESPRIT is identical in the 
high effective SNR regime. To this end, we recall that $R$-D NC Unitary ESPRIT  
includes forward-backward-averaging (FBA) \eqref{fbaa} as well as the transformation into the 
real-valued domain \eqref{nctrans1} as preprocessing steps. We first investigate the effect 
of FBA and state the following theorem:
\begin{thm}
Applying FBA to $\nc{\bm X}$ does not improve the signal subspace estimate.
\label{thm:fba}
\end{thm}
\begin{IEEEproof}
To show this result, we simply use the FBA-processed augmented measurement matrix $\ncfba{\bm X}$ in 
\eqref{fbanc} and compute the Gram matrix $\bm G=\ncfba{\bm X} \ncfbaH{\bm X}$, which yields
\begin{align}\notag
\bm G &=\begin{bmatrix} \nc{\bm X} & \nc{\bm X}\bm \Pi_N  \end{bmatrix} 
\begin{bmatrix} \nc{\bm X} & \nc{\bm X}\bm \Pi_N  \end{bmatrix}^\herm \\
&= 2 \cdot \nc{\bm X} \ncH{\bm X}.
\label{gram}
\end{align}
Thus, the matrix $\bm G$ reduces to the Gram matrix of $\nc{\bm X}$ and the column space of $\nc{\bm X}$ 
is the same as the column space of the Gram matrix of $\nc{\bm X}$. Consequently, FBA has no effect on 
$\nc{\bm X}$. This completes the proof. 
\end{IEEEproof}
Next, we analyze the real-valued transformation as the second preprocessing step of $R$-D NC Unitary 
ESPRIT and formulate the theorem:
\begin{thm}
$R$-D NC Unitary ESPRIT and $R$-D NC Standard ESPRIT with FBA preprocessing perform asymptotically 
identical in the high effective SNR.
\label{thm:realtrans}
\end{thm}
\begin{IEEEproof} 
See Appendix \ref{app:realtrans}.
\end{IEEEproof}
As a result of Theorem \ref{thm:fba} and Theorem \ref{thm:realtrans}, we can conclude that the asymptotic performance 
of $R$-D NC Standard ESPRIT and $R$-D NC Unitary ESPRIT is asymptotically identical in the high effective SNR. 
\section{Single Source Case}
\label{sec:single}
So far, we have derived an MSE expression for both $R$-D NC Standard ESPRIT and $R$-D NC Unitary ESPRIT \eqref{mse}, 
which is deterministic and no Monte-Carlo simulations are required. However, this is only the first step as the 
derived MSE expression is formulated in terms of the subspaces of the unperturbed measurement matrix and hence, 
provides no explicit insights into the influence of the physical parameters, e.g., the SNR, the number of sensors, 
the sample size, etc. Knowing how the performance scales with these system parameters as a second step can facilitate 
array design decisions on the number of required sensors to achieve a certain performance for a specific SNR. Moreover, 
different parameter estimators can be objectively compared to find the best one for particular scenarios. Establishing 
a general formulation for an arbitrary number of sources is an intricate task given the complex dependence of the 
subspaces on the physical parameters. However, special cases can be considered to gain more insights by such an analytical 
performance assessment. Inspired by \cite{roemer2012perf}, we present results for the $R$-D case of a single strictly 
SO non-circular source in this section. To this end, we assume an $R$-D uniform sampling grid, i.e., a ULA in each mode, 
and circularly symmetric white noise. Furthermore, we obtain the same asymptotic estimation error for $R$-D NC 
Standard ESPRIT and $R$-D NC Unitary ESPRIT as proven in the previous section. We also provide results on the single source 
case for the deterministic $R$-D NC CRB \cite{roemer2007nccrb}, which enables the computation of the 
asymptotic efficiency of $R$-D NC Standard ESPRIT and $R$-D NC Unitary ESPRIT for arbitrary dimensions $R$ in closed-form. As an 
example, we compute the asymptotic efficiency for $R=1$.
\subsection{R-D NC Standard ESPRIT and R-D NC Unitary ESPRIT}
As the asymptotic performance of both algorithms is the same, it is again sufficient to simplify the MSE expression in \eqref{mse}
for $R$-D NC Standard ESPRIT. We have the following result:
\begin{thm}
For the case of an $M$-element $R$-D uniform sampling grid with an $M_r$-element ULA in the $r$-th mode, a single strictly 
non-circular source ($d=1$), and circularly symmetric white noise, the MSE of $R$-D NC Standard and $R$-D NC Unitary ESPRIT 
in the $r$-th mode is given by
\begin{equation}
\mathbb E\left\{(\Delta\rd{\mu})^2\right\} = \frac{1}{\hat{\rho}}\cdot\frac{M_r}{M(M_r-1)^2} + \mathcal O\left\{\frac{1}{\hat{\rho}^2}\right\},
\label{ncunit_single}
\end{equation}
where $\hat{\rho}$ represents the effective SNR $\hat{\rho}=N\hat{P}_\mathrm{s}/\sigma_\mathrm{n}^2$ with $\hat{P}_\mathrm{s}$ being 
the empirical source power given by $\hat{P}_\mathrm{s}=\twonorm{\bm s}^2/N$ and $\bm s \in \compl^{N \times 1}$. 
\label{thm:ncunit_single}
\end{thm}
\begin{IEEEproof}
See Appendix \ref{app:ncunit_single}.
\end{IEEEproof}
In a similar fashion, it can be shown that for $R$-D NC Unitary ESPRIT, we arrive at the same MSE result as in 
\eqref{ncunit_single}. Moreover, the expression \eqref{ncunit_single} is equivalent to the ones obtained in 
\cite{roemer2012perf} for the non-NC counterparts. Thus, no improvement in terms of the estimation accuracy 
can be achieved by applying $R$-D NC Standard ESPRIT or $R$-D NC Unitary ESPRIT for a single strictly non-circular 
source. This can also be seen from the result \eqref{nccrb_single} for the deterministic $R$-D NC CRB provided in the next 
subsection, which is also the same as in the non-NC case \cite{roemer2012perf}.
\subsection{Deterministic R-D NC Cram\'{e}r-Rao Bound}
In this part, we simplify the $R$-D extension of the deterministic 2-D NC Cram\'{e}r-Rao Bound derived in 
\cite{roemer2007nccrb} for the special case of a single strictly non-circular source. The result is shown 
in the next theorem:
\begin{thm}
For the case of an $M$-element $R$-D uniform sampling grid with an $M_r$-element ULA in the $r$-th mode and a single strictly 
non-circular source ($d=1$), the deterministic $R$-D NC Cram\'{e}r-Rao Bound can be simplified to
\begin{align}
\nc{\bm C} = \mathrm{diag}\Big\{\big[C^{{\rm (nc)}(1)},\ldots,C^{{\rm (nc)}(R)}\big]^\trans\Big\}, \label{nccrb_single}
\end{align}
where \begin{center}$\displaystyle C^{{\rm (nc)}(r)} = \frac{1}{\hat{\rho}} \cdot \frac{6}{M (M^2_r-1)}$.\end{center} 
\label{thm:nccrb}
\end{thm}
\vspace{0.2em}
\begin{IEEEproof}
See Appendix \ref{app:nccrb}.
\end{IEEEproof}
Using the expressions \eqref{ncunit_single} and \eqref{nccrb_single}, we can analytically compute the asymptotic 
efficiency of the proposed algorithms for arbitrary dimensions $R$. The result for $R=1$ is given in the next 
subsection.
\subsection{Asymptotic Efficiency of 1-D NC Standard and 1-D NC Unitary ESPRIT}
Under the stated assumptions, the asymptotic efficiency for the 1-D case of NC Standard ESPRIT and NC Unitary 
ESPRIT, where $M_r=M$, can be explicitly computed as 
\begin{equation}
\eta=\lim_{\hat{\rho} \to \infty} \frac{\nc{C}}{\mathbb E\{(\Delta\mu)^2\}}=\frac{6(M-1)}{M(M+1)}.
\label{asyeff_single}
\end{equation}
Again, the 1-D asymptotic efficiency \eqref{asyeff_single} is equivalent to the one derived in 
\cite{roemer2012perf}, i.e., no gains are obtained from non-circular sources. It should be noted 
that $\eta$ is only a function of the array geometry, i.e., the number of sensors $M$. The outcome 
of this result is that 1-D NC ESPRIT-type algorithms using LS are asymptotically efficient for $M=2$ 
and $M=3$ for a single source. However, they become less efficient when the number of sensors grows, 
in fact, for $M\rightarrow \infty$ we have $\eta \rightarrow 0$. A possible explanation could be that 
an $M$-element ULA offers not only the single shift invariance with maximum overlap used in LS, but 
multiple invariances that are not exploited by LS.
\section{Simulation Results}
\label{sec:simulations}
In this section, we provide simulation results to evaluate the performance of the proposed $R$-D NC Standard ESPRIT 
and $R$-D NC Unitary ESPRIT algorithms along with the asymptotic behavior of the presented performance analysis. 
We compare the square root of the analytical MSE expression (``ana'') in \eqref{mse} to the root mean squared 
error (RMSE) of the empirical estimation error (``emp'') of $R$-D NC Standard ESPRIT (NC SE) and $R$-D NC Unitary 
ESPRIT (NC UE) obtained by averaging over 5000 Monte Carlo trials. The RMSE is defined as 
\begin{align}
\mathrm{RMSE}=\sqrt{\frac{1}{Rd}~\mathbb E\left\{\sum_{r=1}^R \sum_{i=1}^d \left(\rd{\mu_i}-\rd{\hat{\mu_i}}\right)^2\right\}},
\end{align}
where $\rd{\hat{\mu_i}}$ is the estimate of $i$-th spatial frequency in the $r$-th mode. Furthermore, we compare 
our results to $R$-D Standard ESPRIT (SE), $R$-D Unitary ESPRIT (UE) as well as the deterministic Cram\'{e}r-Rao bounds for 
circular (Det CRB) and strictly SO non-circular sources (Det NC CRB) \cite{roemer2007nccrb}.
In the simulations, we employ different array configurations consisting of isotropic 
sensor elements with interelement spacing $\delta=\lambda/2$ in all dimensions. The phase reference is chosen to be at the 
centroid of the array. It is assumed for all algorithms that a known number of signals with unit power and symbols $\bm S_0$ 
(cf. Equation \eqref{ncsignal}) drawn from a real-valued Gaussian distribution impinge on the array. Moreover, we assume zero-mean 
circularly symmetric white Gaussian sensor noise according to \eqref{wcsnoise}.
\begin{figure}[t!]
    \centerline{\includegraphics[height=\figheight]{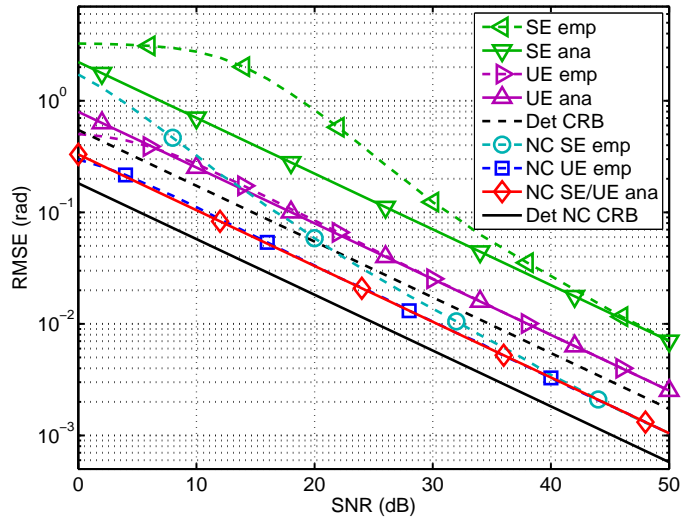}}
    \small
    \caption{Analytical and empirical RMSEs versus SNR for a $4\times 4\times 4$ cubic uniform array ($R=3$), and $N=5$, 
    $d=2$ correlated sources ($\rho=0.9$) at $\mu_1^{(1)}=0$, $\mu_2^{(1)}=0.1$, $\mu_1^{(2)}=0$, $\mu_2^{(2)}=0.1$, 
    $\mu_1^{(3)}=0$, $\mu_2^{(3)}=0.1$ with rotation phases $\varphi_1=0$, $\varphi_2=\pi/2$.}
    \label{fig:rmse_snr}
    \vspace{-1em}
\end{figure}
\begin{figure}[t!]
    \centerline{\includegraphics[height=\figheight]{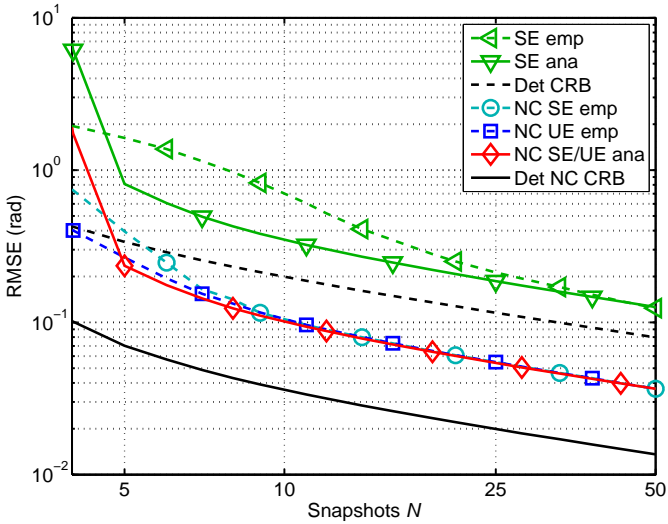}}
    \caption{Analytical and empirical RMSEs versus the snapshots $N$ for the 20-element 2-D array ($R=2$)
    from Fig.~\ref{fig_subsp_2dshinv} and SNR $=10$ dB, $d=3$ uncorrelated sources at 
    $\mu_1^{(1)}=0.25$, $\mu_2^{(1)}=0.5$, $\mu_3^{(1)}=0.75$, $\mu_1^{(2)}=0.25$, $\mu_2^{(2)}=0.5$, 
    $\mu_3^{(2)}=0.75$ with rotation phases $\varphi_1=0$, $\varphi_2=\pi/4$, $\varphi_3=\pi/2$.}
    \label{fig:rmse_snap}
    \vspace{-1em}
\end{figure}

\begin{figure}[b!]
\psfrag{Jxx}{\footnotesize {\color[cmyk]{0,1,1,0} $\ma{\tilde{J}}_1^{(1)}$}}
\psfrag{J12}{\footnotesize {\color[cmyk]{1,0,1,0} $\ma{\tilde{J}}_2^{(1)}$}}
\psfrag{J21}{\footnotesize {\color[cmyk]{0,1,1,0} $\ma{\tilde{J}}_1^{(2)}$}}
\psfrag{J22}{\footnotesize {\color[cmyk]{1,0,1,0} $\ma{\tilde{J}}_2^{(2)}$}}
\begin{center}
\includegraphics[width=0.7\columnwidth]{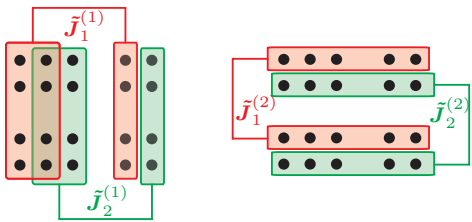}%
\end{center}
\caption{2-D shift invariance for the depicted non-centro-symmetric $5 \times 4$ sampling grid, left: subarrays
for the first (horizontal) dimension, right: subarrays for the second (vertical)
dimension.}%
\label{fig_subsp_2dshinv}%
\end{figure}

\begin{figure}[t!]
    \centerline{\includegraphics[height=\figheight]{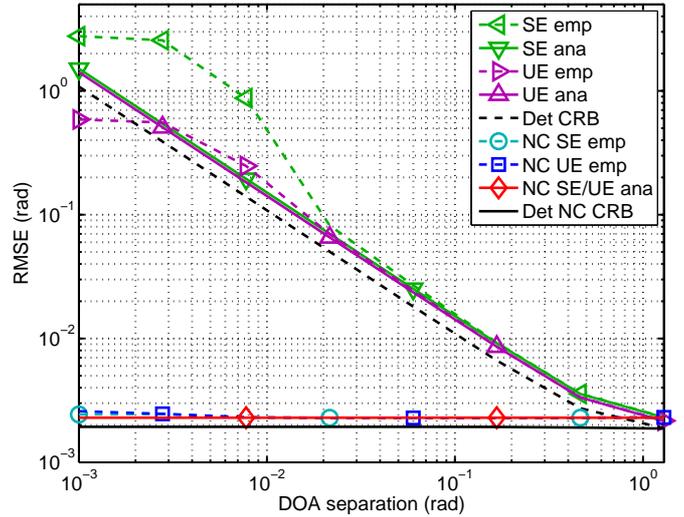}}
    \small
    \caption{Analytical and empirical RMSEs versus the separation (``sep'') of $d=2$ uncorrelated sources at 
    $\mu_1^{(1)}=-\textrm{sep}/2$, $\mu_2^{(1)}=0$, $\mu_1^{(2)}=\textrm{sep}/2$, $\mu_2^{(2)}=\textrm{sep}$ 
    for a $5\times 6$ URA $(R=2)$, $N=5$, $\textrm{SNR}=30$ dB, with rotation phases $\varphi_1=0$, $\varphi_2=\pi/2$ .}
    \label{fig:rmse_musep}
    \vspace{-1em}
\end{figure}
\begin{figure}[t!]
    \centerline{\includegraphics[height=\figheight]{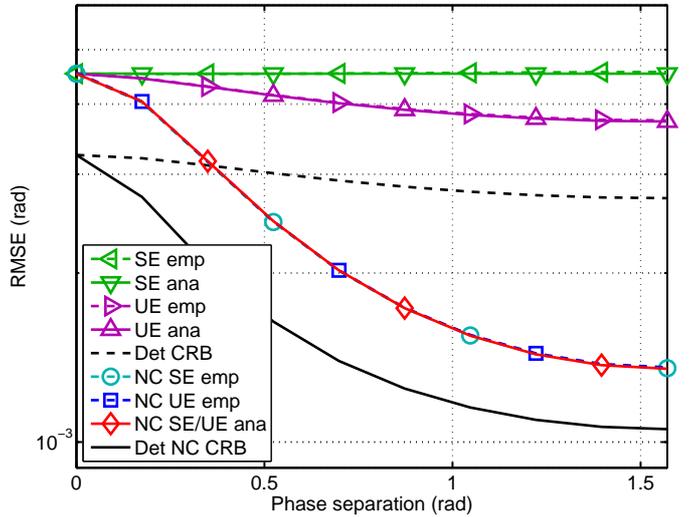}}
    \small
    \caption{Analytical and empirical RMSEs versus the phase separation for a $5\times 6$ URA $(R=2)$, $N=5$, 
    $\textrm{SNR}=30$ dB, $d=2$ uncorrelated sources at $\mu_1^{(1)}=1$, $\mu_2^{(1)}=0.8$, $\mu_1^{(2)}=1$, $\mu_2^{(2)}=0.8$.}
    \label{fig:rmse_phasep}
    \vspace{-1em}
\end{figure}
\begin{figure}[t!]
    \centerline{\includegraphics[height=\figheight]{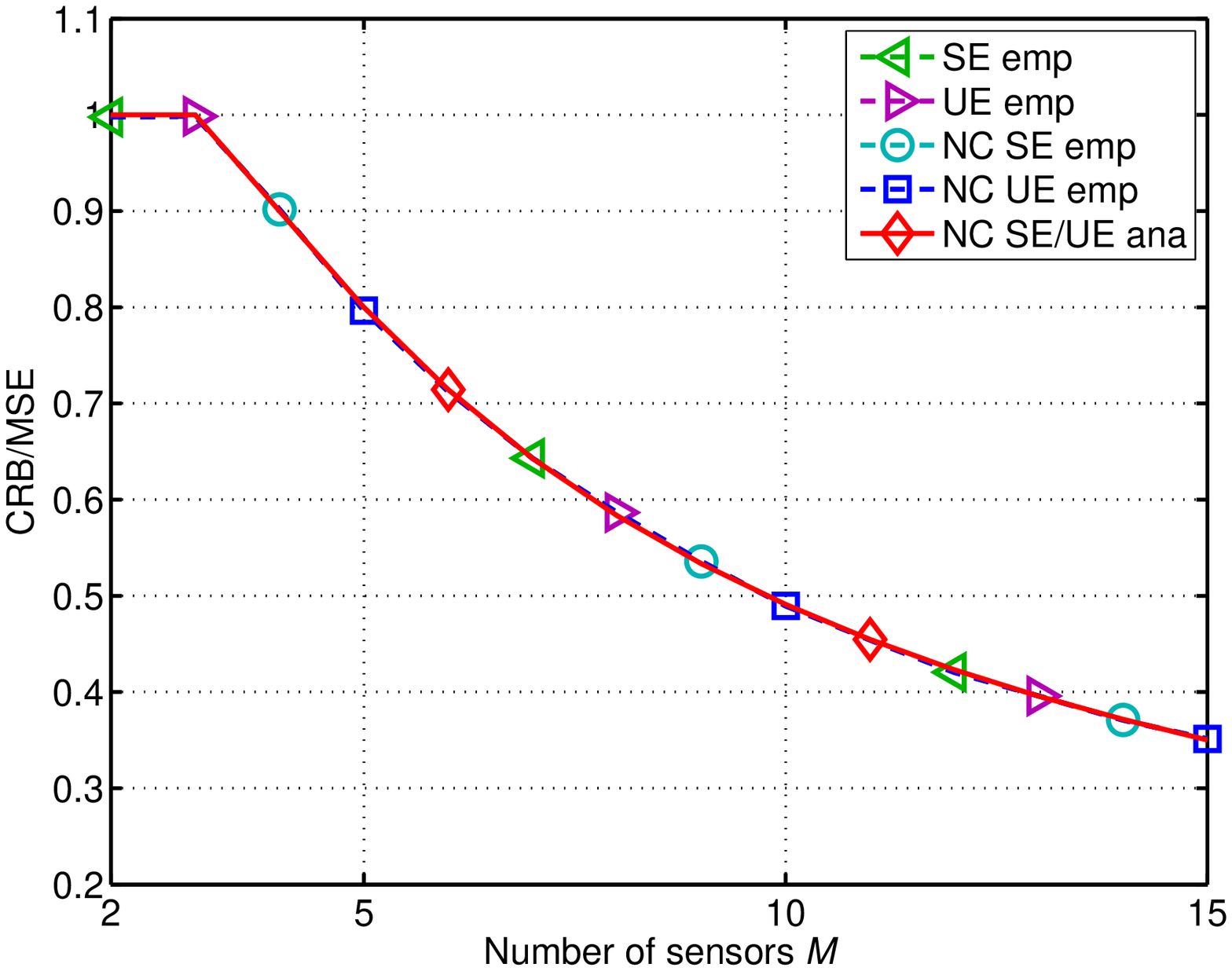}}
    \small
    \caption{Asymptotic efficiency versus $M$ of a ULA $(R=1)$ for a single strictly non-circular source with an effective SNR of $46$ dB 
    ($P_\mathrm{s} = 0$ dB, $N = 4$, $\sigma_n^2 = 10^{-4}$).}
    \label{fig:eff}
    \vspace{-1em}
\end{figure}

Fig.~\ref{fig:rmse_snr} illustrates the RMSE versus the SNR, where we consider a $4 \times 4 \times 4$ uniform cubic 
array with $N=5$ available observations of $d=2$ sources with the spatial frequencies $\mu_1^{(1)}=0$, 
$\mu_2^{(1)}=0.1$, $\mu_1^{(2)}=0$, $\mu_2^{(2)}=0.1$, $\mu_1^{(3)}=0$, and $\mu_2^{(3)}=0.1$, and a real-valued 
pair-wise correlation of $\rho=0.9$. The rotation phases contained in $\bm \Psi$ are given by $\varphi_1=0$ and 
$\varphi_2=\pi/2$. In Fig.~\ref{fig:rmse_snap}, we depict the RMSE versus the number of snapshots $N$ for the 
non-centro-symmetric 2-D array with $M=20$ given in Fig.~\ref{fig_subsp_2dshinv}, where we also provide the subarrays in both 
dimensions. The SNR is fixed at $10$ dB and we have $d=3$ uncorrelated sources with the spatial frequencies 
$\mu_1^{(1)}=0.25$, $\mu_2^{(1)}=0.5$, $\mu_3^{(1)}=0.75$, $\mu_1^{(2)}=0.25$, $\mu_2^{(2)}=0.5$, and $\mu_3^{(2)}=0.75$. 
The rotation phases are given by $\varphi_1=0$, $\varphi_2=\pi/4$, and $\varphi_3=\pi/2$. Note that 2-D Unitary 
ESPRIT cannot be applied as the array is not centro-symmetric. It is apparent from Fig.~\ref{fig:rmse_snr} and 
Fig.~\ref{fig:rmse_snap} that in general, the NC schemes perform better than their non-NC counterparts. 
Specifically, $R$-D NC Unitary ESPRIT provides a lower estimation error than $R$-D NC Standard ESPRIT 
for low SNRs and a low sample size. Moreover, the analytical results agree well with the empirical estimation errors for 
high effective SNRs, i.e., when either the SNR or the number of samples becomes large. This also validates that the asymptotic 
performance of $R$-D NC Standard ESPRIT and $R$-D NC Unitary ESPRIT is identical as both coincide with the analytical curve. 
Note that the performance of the proposed algorithms can degrade if the signals' non-circularity is not perfectly strict.

In Fig.~\ref{fig:rmse_musep}, we show the RMSE as a function of the separation (``sep'') between $d=2$ uncorrelated sources 
located at $\mu_1^{(1)}=-\textrm{sep}/2$, $\mu_2^{(1)}=0$, $\mu_1^{(2)}=\textrm{sep}/2$, $\mu_2^{(2)}=\textrm{sep}$ with 
the rotation phases $\varphi_1=0$, $\varphi_2=\pi/2$. We employ a $5 \times 6$ uniform rectangular array (URA), $N=5$ snapshots, 
and the SNR is fixed at $30$ dB. 
Fig.~\ref{fig:rmse_phasep} demonstrates the RMSE as a function of the non-circularity phase separation $\Delta\varphi$
of the $d=2$ uncorrelated sources with the spatial frequencies $\mu_1^{(1)}=1$, $\mu_2^{(1)}=0.8$, $\mu_1^{(2)}=1$, and 
$\mu_2^{(2)}=0.8$. The remaining parameters are kept the same. Again, it can be seen from Fig.~\ref{fig:rmse_musep} and 
Fig.~\ref{fig:rmse_phasep} that the analytical results match the empirical ones. But more importantly, the 
gain of the NC ESPRIT-type methods increases if the sources approach each other. Furthermore, as a substantial 
feature of strictly non-circular sources, it is observed that for two 
uncorrelated sources with a phase separation of $\Delta\varphi=\pi/2$, the sources entirely decouple as if each 
of them was present alone. In this case, the achievable gain from strictly non-circular sources is largest, which is 
verified by Fig.~\ref{fig:rmse_phasep}. This decoupling effect was also shown analytically for the Det NC CRB in 
\cite{roemer2007nccrb} and recently for NC Standard ESPRIT in \cite{steinwandt2014twosource}.

In the final simulation, we consider the single source case, which was used in Section \ref{sec:single} to express 
the analytical MSE equations of $R$-D NC Standard ESPRIT and $R$-D NC Unitary ESPRIT only in terms of the physical parameters, 
i.e., the array size $M$ and the effective SNR. Fig.~\ref{fig:eff} shows the asymptotic efficiency \eqref{asyeff_single} for the 
case $R=1$ versus the number of sensors $M$ of a ULA. The effective SNR is set to $46$ dB, where $P_\mathrm{s} = 0$ dB, 
$N = 4$, and $\sigma_n^2= 10^{-4}$. This plot validates the fact that 1-D NC Standard ESPRIT and 1-D NC Unitary 
ESPRIT using LS become increasingly inefficient for $M>3$. It should be stressed that the same curves are obtained for 1-D 
Standard ESPRIT and 1-D Unitary ESPRIT. Hence, no gain is achieved from a single strictly non-circular source.
\section{Conclusion}
\label{sec:conclusions}
In this paper, we have presented the $R$-D NC Standard ESPRIT and $R$-D NC Unitary ESPRIT parameter estimation 
algorithms specifically designed for strictly SO non-circular sources and shift-invariant arrays that are not 
necessarily centro-symmetric. We have also derived a first-order analytical performance analysis of both 
algorithms. Our results are based on a first-order expansion of the estimation error in terms of the 
explicit noise perturbation, which is required to be small compared to the signals but no assumptions 
about the noise statistics are needed. We have also derived MSE expressions that only depend on the 
finite SO moments of the noise and merely assume the noise to be zero-mean. All the resulting expressions 
are asymptotic in the effective SNR, i.e., they become accurate for either high SNRs or a large sample 
size. Furthermore, we have analytically proven that $R$-D NC Standard ESPRIT and $R$-D NC Unitary ESPRIT have 
the same asymptotic performance in the high effective SNR regime. However, $R$-D NC Unitary ESPRIT 
should be preferred due to its real-valued operations and its better performance at low effective 
SNRs. We have also computed the 1-D asymptotic efficiency for a single source and found that no gain 
from non-circular sources is achieved in this case. Simulations demonstrate that 
for more than one strictly non-circular source, the NC gain is largest for closely-spaced sources and 
a rotation phase separation of $\pi/2$. 
\appendices
\section{Proof of Theorem 1} 
\label{app:shift}
We consider the 1-D case for simplicity and start by inserting $\nc{\bm J}_1$ and $\nc{\bm J}_2$ into 
\eqref{shiftnc}, which yields

\vspace{-1em}
\footnotesize
\begin{align}
\begin{bmatrix} \bm{J}_1 \bm{A} \notag \\ 
\bm{\Pi}_{\subsel{M}} \bm{J}_2 \bm{\Pi}_M \bm{\Pi}_M \bm{A}^\conj \bm{\Psi}^\conj \bm{\Psi}^\conj
\end{bmatrix} 
\bm{\Phi} 
& = \begin{bmatrix} \bm{J}_2 \bm{A} \\ 
\bm{\Pi}_{\subsel{M}} \bm{J}_1 \bm{\Pi}_M \bm{\Pi}_M \bm{A}^\conj \bm{\Psi}^\conj \bm{\Psi}^\conj
\end{bmatrix}\!\!. \label{eqn_app_shift}
\end{align}
\normalsize
The first $\subsel{M}$ rows are given by $\bm{J}_1\bm{A}\bm{\Phi} = \bm{J}_2
\bm{A}$, which was assumed for the theorem. The second $\subsel{M}$ rows can be simplified by multiplying 
from the left with $\bm{\Pi}_{\subsel{M}}$ and then using the fact that $\bm{\Pi}_M\bm{\Pi}_M = \bm{I}_M$. 
We obtain
\begin{align}
\bm{J}_2 \bm{A}^\conj \bm{\Psi}^\conj \bm{\Psi}^\conj \bm{\Phi}
= \bm{J}_1 \bm{A}^\conj \bm{\Psi}^\conj \bm{\Psi}^\conj.
\end{align}
As $\bm{\Psi}$ and $\bm{\Phi}$ are diagonal, they commute. Then, multiplying twice by 
$\bm{\Psi}$ from the right-hand side cancels $\bm{\Psi}$ as $\bm{\Psi}^\conj\bm{\Psi} = \bm{I}_d$ and we 
are left with
\begin{align}
\bm{J}_2 \bm{A}^\conj \bm{\Phi}
& = \bm{J}_1 \bm{A}^\conj \notag \\
\bm{J}_2 \bm{A}^\conj 
& = \bm{J}_1 \bm{A}^\conj \bm{\Phi}^\conj, 
\label{eqn_app_shift2}
\end{align}
where in the last step we have multiplied with $\bm{\Phi}^\conj$ from the right-hand side and used the fact 
that $\bm{\Phi}^\conj\bm{\Phi} = \bm{I}_d$.\footnote{This equality only holds in the assumed case of undamped 
exponentials (cf. the model in \eqref{rdmodel}), where the spatial frequencies $\rd{\mu}_i$ are real.} Finally, 
conjugating \eqref{eqn_app_shift2} shows that this expression  
is equivalent to $\bm{J}_1\bm{A} \bm{\Phi} = \bm{J}_2 \bm{A}$, which was again assumed for the theorem. This 
concludes the proof. \qed
\vspace{-1em}
\section{Proof of Equation \eqref{nctrans}} 
\label{app:trans}
The real-valued transformation is carried out using sparse left $\bm \Pi$-real matrices of even order according 
to \eqref{leftq}. Expanding \eqref{nctrans1} yields
\begin{align}
&\varphi(\ncfba{\bm X}) = \bm Q^\herm_{2M} \ncfba{\bm X} \bm Q_{2N} \notag\\
&=\frac{1}{2} \cdot \begin{bmatrix} \bm I_M & \bm \Pi_M \\ -\j\bm I_M & \j\bm \Pi_M \end{bmatrix}
\begin{bmatrix} \nc{\bm X} & \nc{\bm X} \bm \Pi_N  \end{bmatrix}
\begin{bmatrix} \bm I_N & \j\bm I_N \\ \bm \Pi_N & -\j\bm \Pi_N \end{bmatrix}\notag\\
&=\begin{bmatrix} \bm I_M & \bm \Pi_M \\ -\j\bm I_M & \j\bm \Pi_M \end{bmatrix}
\begin{bmatrix} \bm X & \bm 0_{M\times N} \\ \bm \Pi_M\bm X^\conj & \bm 0_{M\times N}\end{bmatrix}\notag\\
&=\begin{bmatrix} \bm X + \bm X^\conj & \bm 0_{M\times N} \\ -\j\bm X + \j\bm X^\conj & \bm 0_{M\times N}\end{bmatrix}
= 2\cdot\begin{bmatrix} \realof{\bm X} & \bm 0_{M\times N}\\ \imagof{\bm X} & \bm 0_{M\times N} \end{bmatrix}, \notag
\end{align}
where we have used the fact that $-\j x + \j x^\conj = 2\cdot\imagof{x} ~\forall x\in\mathbb C$.
This completes the proof. \qed
\vspace{-0.5em}
\section{Proof of Theorem \ref{thm:realtrans}}
\label{app:realtrans}
For simplicity, we only present the proof for the 1-D case, but the approach adopted here carries 
over to the $R$-D case straightforwardly. The estimated parameters after the real-valued transformation (NC Unitary ESPRIT) 
are extracted in a different manner as in the forward-backward-averaged complex-valued case (NC Standard ESPRIT with FBA), i.e. 
using the arctangent function. Hence, we develop a first-order perturbation expansion for the 
real-valued shift invariance equations and then show the equivalence of both cases. To this end, let $\ncfba{\bm X}_0 
\in\compl^{2M \times 2N}$ be the noise-free forward-backward averaged measurement matrix defined by 
decomposing \eqref{fbanc} according to
\begin{align}
\ncfba{\bm X}&=\begin{bmatrix} \nc{\bm X_0} & \nc{\bm X_0}\bm \Pi_N \end{bmatrix} + 
\begin{bmatrix} \nc{\bm N} & \nc{\bm N}\bm \Pi_N \end{bmatrix} \notag\\
&= \ncfba{\bm X}_0 + \ncfba{\bm N}.
\label{fbanc0}
\end{align}
Its SVD can be expressed as
\begin{align}\notag
\ncfba{\bm X}_0 &= \begin{bmatrix} \ncfba{\bm U}_{\rm s} & \ncfba{\bm U}_{\rm n} \end{bmatrix}
\begin{bmatrix} \ncfba{\bm \Sigma}_{\rm s} & \bm 0 \\ \bm 0 & \bm 0 \end{bmatrix} 
\begin{bmatrix} \ncfba{\bm V}_{\rm s} & \ncfba{\bm V}_{\rm n} \end{bmatrix}^\herm,
\end{align}
such that the complex-valued shift invariance equation for the forward-backward-averaged data has the form
\begin{align}
\nc{\bm J}_1 \ncfba{\bm U}_{\rm s} \bm \Gamma = \nc{\bm J}_2 \ncfba{\bm U}_{\rm s},
\label{eqn_app_proof_perf_ue_siecv}
\end{align}
where $\bm \Gamma = \fba{\bm Q} \bm \Lambda \fbainv{\bm Q}$ and $\bm \Lambda = \diagof{\begin{bmatrix}
\lambda_1,\ldots,\lambda_d\end{bmatrix}}$ with $\lambda_i = \expof{\j \mu_i}$, $i=1, 2, \ldots, d$.
Performing the same steps as in Section \ref{sec:SEperf}, the first-order approximation of the estimation 
error after the application of FBA is given by 
\begin{eqnarray}
\begin{aligned}
\Delta\mu_i&=\imagof{\fbaT{\bm p}_i \left(\nc{\bm J_1} \ncfba{\bm U}_{\rm s}\right)^+
\left[\nc{\bm J_2}/\lambda_i \right.\right.\\
&\qquad~~\qquad \left.\left.-\nc{\bm J_1}\right]\Delta \ncfba{\bm U}_{\rm s}\fba{\bm q}_i} + \mathcal O\{\nu^2\},
\label{estpertfba}
\end{aligned}
\end{eqnarray}
where we have simply replaced the corresponding quantities in \eqref{estpert} by their FBA versions. Next, 
we show that the estimation error expansion for the real-valued case is equivalent to \eqref{estpertfba}.

The 1-D real-valued shift-invariance equation 
\begin{align}
\nc{\bm K}_1 \nc{\bm E}_{\rm s} \bm{\Upsilon} = \nc{\bm K}_2 \nc{\bm E}_{\rm s}, 
\label{eqn_app_proof_perf_ue_sierv}
\end{align}
where $\bm \Upsilon = \bm V \bm \Omega \bm V^{-1}$ and $\bm \Omega = \diagof{\begin{bmatrix}
\omega_1, \ldots, \omega_d\end{bmatrix}}$ with $\omega_i = \tan(\mu_i/2)$, $i=1, 2, \ldots, d$,
has the same algebraic form as its complex-valued counterpart in \eqref{eqn_app_proof_perf_ue_siecv}. Therefore, 
the same procedure from \cite{vaccaro1993perf} can be applied to develop a first-order perturbation expansion. 
In fact, following the three steps discussed in \cite{vaccaro1993perf}, we find that the perturbation of $\omega_i$ in 
terms of $\bm \Upsilon$	and the perturbation of $\bm \Upsilon$ in terms of the signal subspace estimation error $\Delta \ncfba{\bm U}_{\rm s}$
lead to the same result, where $\nc{\bm J}_1, \nc{\bm J}_2, \nc{\bm U}_{\rm s}$, and $\bm \Gamma$ are consistently
exchanged by $\nc{\bm K}_1, \nc{\bm K}_2, \nc{\bm E}_{\rm s}$, and $\bm \Upsilon$, respectively. Thus, only the 
perturbation of $\mu_i$ in terms of $\omega_i = \tan(\mu_i/2)$ is to be derived.
Therefore, we compute the Taylor series expansion of $\omega_i$, which is given by
\begin{align}
\omega_i + \Delta \omega & \approx \tan(\mu_i/2) + \Delta \mu \left( \frac{\tan^2(\mu_i/2)}{2} + \frac{1}{2}\right) \notag \\
&= \omega_i + \Delta \mu ~\frac{\omega_i^2+1}{2} \quad\textrm{and hence} \notag\\
\Delta \mu & \approx  \Delta \omega \frac{2}{\omega_i^2+1}. 
\label{eqn_app_proof_perf_ue_step1}
\end{align}
Combining \eqref{eqn_app_proof_perf_ue_step1} with the corresponding real-valued expressions
for the perturbations of $\omega_i$ and $\bm \Upsilon$, we obtain
\begin{eqnarray}
\begin{aligned}
\Delta \mu_i &= \bar{\bm p}_i^\trans \left( \nc{\bm K}_1 \nc{\bm E}_{\rm s}\right)^\pinv \left(\nc{\bm K}_2 - \omega_i \nc{\bm K}_1 \right) \\
&\qquad\qquad \cdot\Delta \nc{\bm E}_{\rm s} \bar{\bm q_i} \frac{2}{\omega_i^2+1},
\label{eqn_app_proof_perf_ue_exp1}
\end{aligned}
\end{eqnarray}
where $\bar{\bm q}_i$ is the $i$-th column of $\bm V$ and $\bar{\bm p}_i^\trans$ is the $i$-th row of $\bm V^{-1}$. 
Moreover, the perturbation of the real-valued subspace $\nc{\bm E}_{\rm s}$ is expanded in terms 
of the transformed noise contribution $\varphi(\ncfba{\bm N}) = \bm Q_{2M}^\herm \ncfba{\bm N} \bm Q_{2N}$ as
\begin{align}
\Delta \nc{\bm E}_{\rm s}  = \nc{\bm E}_{\rm n} \ncH{\bm E}_{\rm n} \varphi(\ncfba{\bm N}) 
\nc{\bm W}_{\rm s} \bm \Sigma^{(\varphi)^{-1}}_{\rm s}, 
\label{eqn_app_proof_perf_ue_expsub}
\end{align}
where the required subspaces are obtained from the SVD of the transformed real-valued measurement matrix 
$\varphi(\ncfba{\bm X}_0) = \bm Q_{2M}^\herm \ncfba{\bm X}_0 \bm Q_{2N}  \in\real^{2M \times 2N}$ expressed as
\begin{align}
\varphi(\ncfba{\bm X}_0) & = \begin{bmatrix} \nc{\bm E}_{\rm s} &\!\! \nc{\bm E}_{\rm n} \end{bmatrix}
\begin{bmatrix} \bm \Sigma_{{\rm s}}^{(\varphi)} & \bm 0 \\ \bm 0 & \bm 0 \end{bmatrix}
\begin{bmatrix} \nc{\bm W}_{\rm s} &\!\! \nc{\bm W}_{\rm n} \end{bmatrix}^\herm. \notag
\end{align}
To simplify \eqref{eqn_app_proof_perf_ue_exp1}, it is easy to see that due to the fact that the matrices $\bm Q_p$ 
are unitary, the subspaces of $\varphi(\ncfba{\bm X}_0)$ are also given by choosing
\begin{align}
\nc{\bm E}_{\rm s} & = \bm Q_{2M}^\herm \ncfba{\bm U}_{\rm s}, ~ 
\nc{\bm E}_{\rm n} = \bm Q_{2M}^\herm \ncfba{\bm U}_{\rm n}, ~
\bm \Sigma_{{\rm s}}^{(\varphi)} = \ncfba{\bm \Sigma}_{\rm s} \notag \\
\nc{\bm W}_{\rm s} & = \bm Q_{2N}^\herm \ncfba{\bm V}_{\rm s}, ~
\nc{\bm W}_{\rm n}  = \bm Q_{2N}^\herm \ncfba{\bm V}_{\rm n}. \label{eqn_app_proof_perf_ue_relsvds}
\end{align}
Moreover, the transformed selection matrices $\nc{\bm K}_1$ and $\nc{\ma K}_2$ defined in \eqref{realsel1} 
and \eqref{realsel2} can be reformulated as
\begin{align}
\nc{\ma K}_1 & = ~\bm Q^\herm_{2\subsel{M}} \big(\nc{\bm J}_1 + \nc{\bm J}_2\big) ~\bm Q_{2M} \label{eqn_app_proof_perf_ue_K1} \\
\nc{\bm K}_2 & = \j \cdot \bm Q^\herm_{2\subsel{M}} \big(\nc{\bm J}_1 - \nc{\bm J}_2\big)~ \bm Q_{2M}, \label{eqn_app_proof_perf_ue_K2}
\end{align}
which follows from expanding the real part and the imaginary part according to $2 \cdot \realof{x} = x + x^\conj$ and 
$2 \cdot \imagof{x} = -\j x + \j x^\conj$. The conjugated term $\bm Q_{2\subsel{M}}^\trans \nc{\bm J}_2\bm Q_{2M}^\conj$ 
can be simplified to $\bm Q_{2\subsel{M}}^\herm \nc{\bm J}_1 \bm Q^{}_{2M}$ using the fact that $\nc{\bm J}_1 = \bm 
\Pi_{2\subsel{M}} \nc{\bm J}_2 \bm \Pi^{}_{2M}$ holds since the virtual array is always centro-symmetric as shown in 
Theorem \ref{thm:centro} and the fact that $\bm Q_p$ is left-$\bm \Pi$-real.

Inserting \eqref{eqn_app_proof_perf_ue_expsub} into \eqref{eqn_app_proof_perf_ue_exp1} and applying the identities 
\eqref{eqn_app_proof_perf_ue_relsvds}-\eqref{eqn_app_proof_perf_ue_K2}, we have 
\begin{align}
\Delta \mu_i &= \bar{\bm p}_i^\trans \Big( \big(\nc{\bm J}_1 + \nc{\bm J}_2\big) \ncfba{\bm U}_{\rm s}\Big)^\pinv
\Big( \j \cdot \big(\nc{\bm J}_1 - \nc{\bm J}_2\big) \notag\\
&~~~ - \omega_i \big(\nc{\bm J}_1 + \nc{\bm J}_2\big)\Big) \Delta \ncfba{\bm U}_{\rm s} \bar{\bm q}_i \frac{2}{\omega_i^2+1},    
\label{eqn_app_proof_perf_ue_expsub_ir1}
\end{align}
where $\Delta \ncfba{\bm U}_{\rm s} = \ncfba{\bm U}_{\rm n} \ncfbaH{\ma{U}_{\rm n}} \ncfba{\bm N} \ncfba{\bm V}_s 
\ncfbainv{\bm \Sigma}_{\rm s}$.

In order to further simplify \eqref{eqn_app_proof_perf_ue_expsub_ir1}, we require the following two lemmas:
\begin{lem} 
\label{lem_app_proof_perf_ue_ident}
The following identities are satisfied
\begin{align}
\big(\nc{\bm J}_1 + \nc{\bm J}_2\big) \ncfba{\bm U}_{\rm s} &= \nc{\bm J}_1 \ncfba{\bm U}_{\rm s} \breve{\bm \Gamma} 
\label{eqn_app_proof_perf_ue_j1pj2us}\\
\big(\nc{\bm J}_1 - \nc{\bm J}_2\big) \ncfba{\bm U}_{\rm s} &= \nc{\bm J}_2 \ncfba{\bm U}_{\rm s} \mathring{\bm \Gamma},
\end{align}
where $\breve{\bm \Gamma} = \bm I_d + \bm \Gamma = \fba{\bm Q} \left( \bm I_d + \bm \Lambda\right) \fbainv{\bm Q}$
and $\mathring{\bm \Gamma} = -\bm I_d + \bm \Gamma^{-1} = \fba{\bm Q} \left( -\bm I_d + \bm \Lambda^{-1}\right) \fbainv{\bm Q}$.
\end{lem}
\begin{IEEEproof}
These identities follow straightforwardly from $\nc{\bm J}_1 \ncfba{\bm U}_{\rm s} \bm \Gamma = \nc{\bm J}_2 \ncfba{\bm U}_{\rm s}$ 
by adding $\nc{\bm J}_1 \ncfba{\bm U}_{\rm s}$ to both sides of the equation for the first identity, and subtracting $\nc{\bm J}_1 
\ncfba{\bm U}_{\rm s}$ and substituting $\nc{\bm J}_1 \ncfba{\bm U}_{\rm s}$ by $\nc{\bm J}_2 \ncfba{\bm U}_{\rm s} \bm \Gamma^{-1}$ 
for the second identity.
\end{IEEEproof}
\begin{lem} 
\label{lem_app_proof_perf_ue_upspsi}
In the noiseless case, the solution $\bm \Gamma$ to \eqref{eqn_app_proof_perf_ue_siecv} and the solution $\bm \Upsilon$ to 
\eqref{eqn_app_proof_perf_ue_sierv} have the same eigenvectors, i.e., $\fba{\bm Q} = \bm V$. Moreover, 
their eigenvalues are related as $\omega_i = \j \cdot \frac{1-\lambda_i}{1+\lambda_i}$.
\end{lem}
\begin{IEEEproof}
Starting from $\bm \Upsilon = \big(\nc{\bm K}_1 \nc{\bm E}_{\rm s}\big)^\pinv \nc{\bm K}_2 \nc{\bm E}_{\rm s}$
and replacing $\nc{\bm E}_{\rm s}$ with \eqref{eqn_app_proof_perf_ue_relsvds} and $\nc{\bm K}_n$ with 
\eqref{eqn_app_proof_perf_ue_K1} and \eqref{eqn_app_proof_perf_ue_K2},
we get
\begin{align}
\bm \Upsilon & =  \left(  \big(\nc{\bm J}_1 + \nc{\bm J}_2\big) \ncfba{\bm U}_{\rm s}\right)^\pinv \j \cdot 
\big(\nc{\bm J}_1 - \nc{\bm J}_2\big) \ncfba{\bm U}_{\rm s} \notag \\
& = \j \cdot \breve{\bm \Gamma}^{-1} \bm \Gamma \mathring{\bm \Gamma} = \j \cdot \fba{\bm Q} \left(\bm I_d + \bm \Lambda\right)^{-1} 
\left(\bm I_d - \bm \Lambda\right) \fbainv{\bm Q} \notag \\
& = \fba{\bm Q} \bm \Omega \fbainv{\bm Q},
\end{align}
where $\bm \Omega=\diagof{\j \cdot \big[\frac{1-\lambda_i}{1+\lambda_i}\big]}_{i=1}^d$ and we have used Lemma 
\ref{lem_app_proof_perf_ue_ident} in the first step.
\end{IEEEproof}

Next, we consider the term $\big(\j \cdot \big(\nc{\bm J}_1 - \nc{\bm J}_2\big) - \omega_i \big(\nc{\bm J}_1 + \nc{\bm J}_2\big)\big)$ 
in \eqref{eqn_app_proof_perf_ue_expsub_ir1} and apply the relation $\omega_i = \j\cdot \frac{1-\lambda_i}{1+\lambda_i}$
from Lemma \ref{lem_app_proof_perf_ue_upspsi}. We can then rewrite this term as
$\j \cdot \big(\nc{\bm J}_1 \lambda_i - \nc{\bm J}_2\big) \frac{2}{1+\lambda_i}$. Moreover, the term $\frac{2}{\omega_i^2+1}$ in \eqref{eqn_app_proof_perf_ue_expsub_ir1} can be expressed 
in terms of $\lambda_i$ as $\frac{2}{\omega_i^2+1} 
= \frac{(\lambda_i+1)^2}{2\lambda_i}$. Inserting these relations into \eqref{eqn_app_proof_perf_ue_expsub_ir1}, 
replacing $\big(\nc{\bm J}_1 + \nc{\bm J}_2\big) \ncfba{\bm U}_{\rm s}$ via \eqref{eqn_app_proof_perf_ue_j1pj2us}, 
and substituting $\bar{\bm p}_i=\fba{\bm p}_i$ and $\bar{\bm q}_i=\fba{\bm q}_i$ using Lemma \ref{lem_app_proof_perf_ue_upspsi}, 
yields
\begin{align}
\Delta \mu_i & =  \j \cdot \fbaT{\bm p}_i \breve{\bm \Gamma}^{-1} \left(\nc{\bm J}_1 \ncfba{\bm U}_{\rm s}\right)^\pinv 
\left(\nc{\bm J}_1 \lambda_i - \nc{\bm J}_2\right) \notag\\
&\qquad \cdot\Delta \ncfba{\bm U}_{\rm s} \fba{\bm q}_i \frac{2}{1+\lambda_i} \cdot \frac{(\lambda_i+1)^2}{2\lambda_i} \notag \\
& =  -\j \cdot \fbaT{\bm p}_i \left(\nc{\bm J}_1 \ncfba{\bm U}_{\rm s}\right)^\pinv \left(\nc{\bm J}_2/\lambda_i - \nc{\bm J}_1\right) \notag\\
&\qquad \cdot\Delta \ncfba{\bm U}_{\rm s} \fba{\bm q}_i,
\label{eqn_app_proof_perf_ue_expsub_ir4}
\end{align}
where we used $\fbaT{\bm p}_i \breve{\bm \Gamma}^{-1} = \fbaT{\bm p}_i (1+\lambda_i)^{-1}$ from 
Lemma~\ref{lem_app_proof_perf_ue_ident} in the first equation.

As a final step, we notice that \eqref{eqn_app_proof_perf_ue_expsub_ir4} must be real-valued as we have started 
from the purely real-valued expansion \eqref{eqn_app_proof_perf_ue_exp1} and only used equivalence transforms to 
arrive at \eqref{eqn_app_proof_perf_ue_expsub_ir4}. However, if $-\j z \in\real$ for $z \in\compl$ this implies
that $\realof{z} = 0$ and hence $-\j z = \imagof{z}$. Consequently, \eqref{eqn_app_proof_perf_ue_expsub_ir4} can 
also be written as \eqref{estpertfba} and is therefore equivalent to the first-order expansion for $R$-D 
NC Standard ESPRIT with FBA. This concludes the proof of the theorem. \qed
\vspace{-0.5em}
\section{Proof of Theorem \ref{thm:ncunit_single}} 
\label{app:ncunit_single}
We start the proof by simplifying the MSE expression for $R$-D NC Standard ESPRIT in \eqref{mse}.
In the single source case the noise-free NC measurement matrix can be written as
\begin{align}
\nc{\bm X_0} = \nc{\bm a}(\bm \mu) \bm s^\trans, 
\label{modsingle}
\end{align}
where $\nc{\bm a}(\bm \mu)=[\bm a^\trans(\bm \mu), \tilde{\Psi}\bm \Pi_M \bm a^\herm(\bm \mu)]^\trans 
\in \compl^{2M \times 1}$ is the augmented array steering vector and $\bm a(\bm \mu)=\bm a^{(1)}(\mu^{(1)})
\otimes\cdots\otimes\bm a^{(R)}(\mu^{(R)})~ \in\mathbb C^{M\times 1}$. Moreover, 
$\tilde{\Psi}=\Psi^\conj\Psi^\conj=\expof{-\j2\varphi}$, $\bm s \in \compl^
{N \times 1}$ contains the source symbols, and $\hat{P}_{\rm s} = \twonorm{\bm s}^2 / N$ is the empirical 
source power. In what follows, we drop the dependence of $\nc{\bm a}$ on $\bm \mu$ for notational convenience. 
If we assume a ULA of isotropic elements in each of the $R$ modes, we have $\rd{\bm a} = [1,\expof{\j \rd{\mu}},
\ldots,\expof{\j(M_r-1)\rd{\mu}}]^\trans$ and $\twonorm{\nc{\bm a}}^2 = 2M$. The selection matrices 
$\ncrtil{\bm J}_1$ and $\ncrtil{\bm J}_2$ are then chosen according to \eqref{j1nctil} with $\rd{\bm J_1} = 
[\bm I_{M_r-1},\bm 0_{(M_r-1)\times 1}]$ and $\rd{\bm J_2} = [\bm 0_{(M_r-1)\times 1},\bm I_{M_r-1}]$ 
for maximum overlap, i.e., $\subsel{M_r} = M_r-1$. Note that \eqref{modsingle} is a rank-one matrix and 
we can directly determine the subspaces from the SVD as
\begin{align}
\nc{\bm U_{\rm s}} & = \nc{\bm u_{\rm s}} = \frac{\nc{\bm a}}{\twonorm{\nc{\bm a}}} 
= \frac{\nc{\bm a}}{\sqrt{2M}} \notag\\
\nc{\bm \Sigma_{\rm s}} & = \nc{\sigma_{\rm s}} = \sqrt{2MN\hat{P}_{\rm s}} \notag\\
\nc{\bm V_{\rm s}} & = \nc{\bm v_{\rm s}} = \frac{\bm s^\conj}{\twonorm{\bm s}} 
= \frac{\bm s^\conj}{\sqrt{N \hat{P}_{\rm s}}}. \notag
\end{align}
For the MSE expression in \eqref{mse}, we also require $\bm P^{\perp}_{\nc{\bm a}}=\nc{\bm U_{\rm n}} 
\ncH{\bm U_{\rm n}} = \bm I_{2M} - \frac{1}{2M} \nc{\bm a} \ncH{\bm a}$, which is the projection matrix 
onto the noise subspace. Moreover, we have $\rd{\bm \Gamma}=\expof{\j \rd{\mu}}$ and hence, the eigenvectors are 
$\rd{\bm p_i}=\rd{\bm q_i} = 1$. The SO moments $\nc{\bm R_\mathrm{nn}}$ and $\nc{\bm C_\mathrm{nn}}$ of the noise 
are given by \eqref{wcsnoise}.  
 
Inserting these expressions into \eqref{mse}, we get 

\vspace{-1em}
\small
\begin{eqnarray}
\begin{aligned}
&\quad\expvof{(\Delta \rd{\mu})^2} = \frac{\sigma_{\rm n}^2}{2} \left( \twonorm{\ncrT{\bm r}\nc{\bm W}}^2 \right.\\
&\left. - \realof{\ncrT{\bm r} \nc{\bm W} (\bm I_N\otimes\bm \Pi_{2M}) \left(\ncrT{\bm r} \nc{\bm W}\right)^\trans}\right)
\label{mse_single}
\end{aligned}
\end{eqnarray}
\normalsize
with
\small
\begin{align}
\ncr{\bm r} &= \left[ \left(\ncrtil{\bm J}_1  \frac{\nc{\bm a}}{\sqrt{2M}} \right)^+ \!\!
\left(\ncrtil{\bm J}_2/\expof{\j \rd{\mu}} - \ncrtil{\bm J}_1\right)\right]^\trans, \notag \\ 
\nc{\bm W} & = \left(\frac{1}{\sqrt{2M N \hat{P}_{\rm s}}} \cdot \frac{\bm s^\herm}{\sqrt{N \hat{P}_{\rm s}}}\right) 
\otimes \bm P^{\perp}_{\nc{\bm a}} ~\in\compl^{2M\times 2MN}. \notag 
\end{align}
\normalsize
Note that the term $\ncrT{\bm r}\nc{\bm W}$ can also be written as $\ncrT{\bm r}\nc{\bm W}=\tilde{\bm s}^\trans 
\otimes \rdT{\tilde{\bm a}}$, where

\vspace{-1em}
\small
\begin{align}
\tilde{\bm s}^\trans &= \frac{1}{\sqrt{2M N \hat{P}_{\rm s}}} \cdot \frac{\bm s^\herm}
{\sqrt{N \hat{P}_{\rm s}}}, \notag\\
\rdT{\tilde{\bm a}} &=  \left(\ncrtil{\bm J}_1  \frac{\nc{\bm a}}{\sqrt{2M}} \right)^+\!\!\!\left(\ncrtil{\bm J}_2/
\expof{\j\rd{\mu}} - \ncrtil{\bm J}_1\right)\bm P^{\perp}_{\nc{\bm a}}. \notag 
\end{align}
\normalsize
Thus, after straightforward calculations, the MSE in \eqref{mse_single} is given by
\begin{eqnarray}
\begin{aligned} 
\expvof{(\Delta \rd{\mu})^2} &= \frac{\sigma_{\rm n}^2}{2} \left( \twonorm{\tilde{\bm s}^\trans}^2 \cdot \twonorm{\rdT{\tilde{\bm a}}}^2 \right.\\
&\qquad \bigg. - \realof{\tilde{\bm s}^\trans \tilde{\bm s} \cdot \rdT{\tilde{\bm a}} \bm \Pi_{2M} \rd{\tilde{\bm a}}} \bigg).
\label{msesingle2}
\end{aligned}
\end{eqnarray}
The first term $\twonorm{\tilde{\bm s}^\trans}^2$ of \eqref{msesingle2} can be conveniently 
expressed as $\twonorm{\tilde{\bm s}^\trans}^2 = \frac{1}{2M N\hat{P}_{\rm s}}$. For the second 
term $\big\|\rdT{\tilde{\bm a}}\big\|_2^2$ of \eqref{msesingle2}, we simplify 
$\rdT{\tilde{\bm a}}$ and expand the pseudo-inverse of $\ncrtil{\bm J}_1 \ncr{\bm a}$ using 
the relation $\bm x^\pinv = \bm x^\herm / \twonorm{\bm x}^2$. As $\ncrtil{\bm J}_1$ selects 
$2(M_r-1)$ out of $2M_r$ elements in the $r$-th mode, we have $\big\|\ncrtil{\bm J}_1 
\ncr{\bm a}\big\|_2^2=\frac{M}{M_r}\cdot2(M_r-1)$. Then, taking the shift invariance 
equation $\ncrtil{\bm J}_2 \nc{\bm a} / \expof{\j\rd{\mu}} - \ncrtil{\bm J}_1 \nc{\bm a} = \bm 0$ in the 
$r$-th mode into account, we obtain
\begin{align}
\rdT{\tilde{\bm a}} & = \frac{\sqrt{2M}M_r}{2M(M_r-1)} \left(\ncH{\bm a} \ncrtilH{\bm J}_1 \ncrtil{\bm J}_2/\expof{\j\rd{\mu}} \right. \notag\\
&\qquad \left.- \ncH{\bm a} \ncrtilH{\bm J}_1 \ncrtil{\bm J}_1 \right).
\label{asingle}
\end{align}
As a ULA is centro-symmetric, i.e., \eqref{centro} holds, we can write $\nc{\bm a}=[1,\tilde{\Psi}]^\trans 
\otimes \bm a$. Note that the phase term depending on the phase center in \eqref{centro} cancels 
throughout the derivation and thus has been neglected. Since the vector $\bm a$ and the matrices $\ncrtil{\bm J}_k,~k=1,2$, can 
be written as $\bm a=\bm a^{(1)}\otimes\cdots\otimes\bm a^{(R)}$ and $\ncrtil{\bm J}_k=\bm I_2\otimes \bm I_{\prod_{l=1}^{r-1} M_l} 
\otimes \rd{\bm J}_k \otimes \bm I_{\prod_{l=r+1}^{R} M_l}$, all the unaffected modes can be factored out of 
\eqref{asingle}, yielding 
\begin{align}
&\rdT{\tilde{\bm a}}  = \frac{\sqrt{2M}M_r}{2M(M_r-1)} \cdot\begin{bmatrix} 1 \\ \tilde{\Psi}\end{bmatrix} \otimes \left(\bm a^{(1)}\otimes\cdots\otimes\bm a^{(r-1)} \right)^\herm \notag\\
&\otimes \left(\rdT{\tilde{\bm a}}_1 - \rdT{\tilde{\bm a}}_2 \right) \otimes \left(\bm a^{(r+1)}\otimes\cdots\otimes\bm a^{(R)} \right)^\herm,
\label{asingle2}
\end{align}
where
\begin{align}
\rdT{\tilde{\bm a}}_1 & = \rdH{\bm a} \rdH{\bm J}_1 \rd{\bm J}_2/\expof{\j\rd{\mu}} ~~ \textrm{and} \notag\\
\rdT{\tilde{\bm a}}_2 & = \rdH{\bm a} \rdH{\bm J}_1 \rd{\bm J}_1. \notag
\end{align}
Similarly to \cite{roemer2012perf}, it is easy to verify that
\begin{align}
\rdT{\tilde{\bm a}}_1 &{=} \left[0, \expof{-\j\rd{\mu}},\ldots,\expof{-\j(M_r-2)\rd{\mu}},\expof{-\j(M_r-1)\rd{\mu}}\right] \notag \\
\rdT{\tilde{\bm a}}_2 &{=} \left[1, \expof{-\j\rd{\mu}},\ldots,\expof{-\j(M_r-2)\rd{\mu}},0\right]. \notag
\end{align}
Consequently, we obtain 

\vspace{-1em}
\small
\begin{align}
\twonorm{\rdT{\tilde{\bm a}}}^2 &= \frac{M_r^2}{2M(M_r-1)^2} \cdot 2 \cdot \prod_{n=1}^{r-1} \twonorm{\bm a^{(n)}}^2 \cdot 2 
\cdot \!\! \prod_{n=r+1}^{R} \twonorm{\bm a^{(n)}}^2   \notag \\ 
&= \frac{2M_r^2}{M(M_r-1)^2}\cdot \frac{M}{M_r}=\frac{2M_r}{(M_r-1)^2}.
\label{term2single}
\end{align}
\normalsize
The third term $\tilde{\bm s}^\trans\tilde{\bm s}$ of \eqref{msesingle2} can be simplified as 
$\tilde{\bm s}^\trans\tilde{\bm s} = \frac{\tilde{\Psi}}{2M N\hat{P}_{\rm s}}$, where we have 
used the equality $\bm s = \Psi\bm s_0$ and the fact that $\bm s_0^\trans\bm s_0=N\hat{P}_{\rm s}$.
Moreover, using \eqref{asingle2}, the last term of \eqref{msesingle2} can be reduced to 
$\rdT{\tilde{\bm a}}\bm \Pi_{2M} \rd{\tilde{\bm a}} = -\frac{2M_r\tilde{\Psi}^\conj}{(M_r-1)^2}$. 
Inserting these results into \eqref{msesingle2}, we finally obtain for the MSE of $R$-D NC Standard 
ESPRIT
\begin{align} 
\expvof{(\Delta \rd{\mu})^2} = \frac{\sigma_{\rm n}^2}{N\hat{P}_{\rm s}}\cdot \frac{M_r}{M(M_r-1)^2},
\label{msesingle3}
\end{align}
which is the desired result. \qed
\section{Proof of Theorem \ref{thm:nccrb}} 
\label{app:nccrb}
We first state the expression for the deterministic NC CRB $\nc{\bm C}$ derived in \cite{roemer2007nccrb}, which is given in 
the $R$-D case by
\begin{align}
&\nc{\bm C} = \frac{\sigma^2_\mathrm{n}}{2N} \cdot \realof{\bm J}^\inv 
\end{align}
with
\begin{align}
&\bm J = \left(\bm{G}_2 - \bm{G}_1 \bm{G}_0^{-1} \bm{G}_1^\trans \right) \odot \hat{\bm R}^{(R)} \notag \\ 
&+ \!\left[ \left( \bm{G}_1 \bm{G}_0^{-1} \bm{H}_0 \right) \odot \hat{\bm R}^{(R)} \right]\! \left[ \left( \bm{G}_0 - \bm{H}_0^\trans \bm{G}_0^{-1} \bm{H}_0 
\right) \odot \hat{\bm R}^{(R)} \right]^{-1} \notag\\
& \cdot \left[ \left(\bm{H}_1^\trans -  \bm{H}_0^\trans \bm{G}_0^{-1} \bm{G}_1^\trans \right) \odot \hat{\bm R}^{(R)} \right] 
+ \left[ \bm{H}_1 \odot \hat{\bm R}^{(R)} \right] \notag\\
& \cdot \left[ \bm{G}_0 \odot \hat{\bm R}^{(R)} \right]^{-1} \!\!\!\cdot \left[ \left( \bm{H}_0^\trans \bm{G}_0^{-1} \bm{G}_1^\trans \right) \odot 
\hat{\bm R}^{(R)} \right] + \left[ \bm{H}_1 \odot \hat{\bm R}^{(R)} \right] \notag\\ 
& \cdot \left[ \bm{G}_0 \odot \hat{\bm R}^{(R)} \right]^{-1}
\cdot \left[ \left( \bm{H}_0^\trans \bm{G}_0^{-1} \bm{H}_0 \right) \odot \hat{\bm R}^{(R)} \right] \notag\\
& \cdot \left[ \left( \bm{G}_0 - \bm{H}_0^\trans \bm{G}_0^{-1} \bm{H}_0 \right) \odot \hat{\bm R}^{(R)} \right]^{-1} \!\!\!{\cdot} \left[ \left( \bm{H}_0^\trans \bm{G}_0^{-1} \bm{G}_1^\trans \right) \odot \hat{\bm R}^{(R)} \right] \notag\\
&- \left[ \bm{H}_1 \odot \hat{\bm R}^{(R)} \right] \cdot \left[ \left( \bm{G}_0 - \bm{H}_0^\trans \bm{G}_0^{-1} \bm{H}_0 
\right) \odot \hat{\bm R}^{(R)} \right]^{-1} \notag\\ 
&\cdot \left[ \bm{H}_1^\trans \odot \hat{\bm R}^{(R)} \right],
\label{eqn_thekillercrb}
\end{align}
where $\hat{\bm R}^{(R)} = \bm 1_R \otimes \hat{\bm R}_{S_0}$ and $\hat{\bm R}_{S_0}=\bm S^{}_0\bm S_0^\trans /N$. The matrices $\bm G_n,~\bm H_n,~n=0,1,2$, 
are defined as
\begin{align}
\bm G_0 &= \realof{\bm \Psi^\conj \bm A^\herm \bm A \bm \Psi},~\bm H_0 = \imagof{\bm \Psi^\conj \bm A^\herm \bm A \bm \Psi} \label{g0}\\
\bm G_1 &= \realof{(\bm I_R \otimes \bm \Psi^\conj) \bm D^\herm \bm A \bm \Psi}, \label{g1}\\
\bm H_1 &= \imagof{(\bm I_R \otimes \bm \Psi^\conj) \bm D^\herm \bm A \bm \Psi}, \label{g2}\\
\bm G_2 &= \realof{(\bm I_R \otimes \bm \Psi^\conj) \bm D^\herm \bm D (\bm I_R \otimes \bm \Psi)}, 
\label{g3}
\end{align}
where $\bm D = [\bm D^{(1)},\ldots,\bm D^{(R)}] \in\compl^{M\times dR}$ with $\bm D^{(r)} = [\rd{\bm d}_1,\ldots,
\rd{\bm d}_d] \in\compl^{M\times d},~r=1,\ldots,R$. The vectors $\rd{\bm d}_i,~i=1,\ldots,d$, contain the partial 
derivatives $\partial \bm a(\bm \mu_i)/\partial \rd{\mu}_i$. In the special case $d=1$, the array steering matrix 
$\bm A$ reduces to $\bm a(\bm \mu)$, $\bm D=[\bm d^{(1)},\ldots,\bm d^{(R)}]\in\compl^{M\times R}$, $\bm \Psi=\expof{\j\varphi}$, 
and $\bm \hat{R}_{S_0} = \bm s_0^\trans\bm s_0^{} /N=\hat{P}_\mathrm{s}$, where $\bm s_0 \in\real^{N\times 1}$. Dropping the dependence of $\bm a$ on $\bm \mu$ 
and using the fact that $\bm a=\bm a^{(1)}\otimes\cdots\otimes \bm a^{(R)}$, we obtain
\begin{align}
\rd{\bm d} = \bm a^{(1)}\otimes\cdots\otimes \bm a^{(r-1)} \otimes \rd{\tilde{\bm d}} \otimes \bm a^{(r+1)}\otimes\cdots\otimes \bm a^{(R)}. \notag
\end{align}
For a ULA in each of the $R$ modes, we have $\rd{\bm a} = [1,\expof{\j\rd{\mu}},\ldots,\expof{\j(M_r-1)\rd{\mu}}]^\trans$ 
and $\rd{\tilde{\bm d}} = \partial \rd{\bm a}/\partial \rd{\mu}= \j\cdot[0,\expof{\j\rd{\mu}},\ldots,(M_r-1)~\expof{\j(M_r-1)
\rd{\mu}}]^\trans$. Then, similarly to \cite{roemer2012perf}, the terms $\bm a^\herm \bm a$, $\bm d^{{(r_1)}^\herm} \bm d^{(r_2)}$, 
and $\rdH{\bm d} \bm a$ in \eqref{g0}-\eqref{g3} become $\bm a^\herm \bm a=M$,
\begin{align}
\bm d^{{(r_1)}^\herm} \bm d^{(r_2)} = \begin{cases} \frac{1}{6} M(M_r-1)(2M_r-1) &\mbox{if } r_1=r_2=r \\ 
\frac{1}{4} M(M_{r_1}-1)(M_{r_2}-1) & \mbox{if } r_1 \neq r_2. \end{cases} \notag
\end{align}
and
\begin{align}
\rdH{\bm d} \bm a= -\j\cdot\frac{1}{2}M(M_r-1). \notag
\end{align}
Thus, the terms \eqref{g0}-\eqref{g3} simplify to
\begin{align}
&\!\!\bm G_0 = M, \qquad\bm H_0=\bm G_1= 0, \label{g4}\\
&\!\!\bm H_1 = \bm h_1 \in\real^{R\times 1} \quad \textrm{with}~~[\bm h_1]^{}_{r} = -\frac{1}{2}M(M_r-1), \label{g5}\\
&\!\!\!\![\bm G_2]^{}_{r_1,r_2} {=} \begin{cases} \frac{1}{6} M(M_r-1)(2M_r-1) ~~\!\mbox{if } r_1=r_2=r \\ 
\frac{1}{4} M(M_{r_1}-1)(M_{r_2}-1) ~\mbox{if } r_1 \neq r_2.\label{g6} \end{cases}
\end{align}
After inserting \eqref{g4}-\eqref{g6} into \eqref{eqn_thekillercrb}, we obtain 
\begin{align}
\bm J = \hat{P}_\mathrm{s} \left(\bm G_2 - \frac{1}{M} \bm h_1^{} \bm h_1^\trans \right). 
\end{align}
It can then be verified that $\bm J$ is a real-valued diagonal matrix with the entries 
$[\bm J]^{}_{r,r} = \frac{\hat{P}_\mathrm{s}}{12} \cdot M(M^2_r-1)$ on its diagonal. Finally, $\nc{\bm C}$ is given by
\begin{align}
\!\nc{\bm C} = \frac{\sigma^2_\mathrm{n}}{2N}\cdot \realof{\bm J}^\inv \!= \mathrm{diag}\Big\{\big[C^{{\rm (nc)}(1)},\ldots,C^{{\rm (nc)}(R)}\big]^\trans\Big\},
\notag
\end{align}
where
\begin{align}
C^{{\rm (nc)}(r)}= \frac{\sigma^2_\mathrm{n}}{N\hat{P}_\mathrm{s}} \cdot \frac{6}{M (M^2_r-1)}
\label{app_nccrb_single}
\end{align}
which is the desired result. \qed 
%
%
%
%
\bibliographystyle{IEEEbib}
\bibliography{refs_nc_unitary_esprit}

\end{document}